\pdfoutput=1 
\documentclass[twocolumn]{aa}

\usepackage{mathtools}
\usepackage{braket}
\usepackage{bm}
\usepackage{empheq} 
\usepackage{latexsym}   
\usepackage{amsfonts}   
\usepackage{amstext}    
\usepackage{amsmath,amssymb}
\usepackage{amscd}
\usepackage{color}
\usepackage{multirow}  

\usepackage{txfonts}
\begin{document}

   \title{Formation and morphology of anomalous \\ solar circular polarization}


   \author{E.S. Carlin  \inst{1,2,3}
          }

   \institute{Instituto de Astrofisica de Canarias, E-38205, La
             Laguna, Tenerife, Spain\\
             \email{ecarlin@iac.es}
         \and
             Universidad de La Laguna, Dpto. Astrofisica, E-38206, La
  Laguna, Tenerife, Spain
         \and
         Istituto Ricerche Solari Locarno, 6600, Locarno, Switzerland
             }

   \date{Received January 31, 2019; accepted ...}

 
  \abstract
 {The morphology of spectral line polarization is the most valuable
    observable to investigate the magnetic and dynamic solar
    atmosphere. However, in order to develop solar
    diagnosis, it is fundamental to understand the different kinds of anomalous solar signals 
    that are routinely found in linear and circular  polarization (LP,CP).}
{We aim to explain and characterize the morphology of solar CP
    signals considering NLTE effects.}
{An analytical two-layer model of the polarized radiative transfer
  equation is developed and used to solve the NLTE (Non-Local
    Thermodinamical Equilibrium) problem with atomic
  polarization in a semi-parametric way. 
The potential of the model for reproducing
 anomalous CP is shown with detailed calculations
 and examples. 
A new approach based on the zeroes of polarization signals is developed to explain their morphology.}
{We have obtained a comprehensive model that insightfully describes 
  the formation of solar polarization with
  certain precision without sacrificing key physical ingredients or resorting to complex atmospheric models. The essential physical behavior of
 dichroism and atomic orientation has been described, introducing
the concepts of dichroic inversion, neutral and reinforcing
medium, critical intensity spectrum, and critical source function. 
We show that the zero-crossings of the CP spectrum are useful
to classify its morphology and understand its formation. This led to identification and
explanation of the morphology of the seven most characteristic
CP signals that a single (depth-resolved) scattering layer can produce. 
We find that a minimal number of two magnetic layers along the line of sight is 
required to fully explain anomalous solar CP signals and that the 
 morphology and polarity of Stokes V depends on
 magnetic, radiative, and atomic
``polarities''. Some implications of these results are presented through a preliminary
modeling of anomalous CP signals in the Fe {\sc
  i} $1564.8$ nm and Na {\sc i} D lines.}
{}

   \keywords{Sun: atmosphere --
                radiative transfer --
                polarization --
                lines: formation
               }

   \maketitle
%
\section{Introduction}

The normal antisymmetric spectrum of circular polarization 
in solar spectral lines is due to magnetic fields modifying the optical
properties of the solar plasma through the Zeeman effect. 
The frequent variations of antisymmetric signals into asymmetric ones have
been well measured and studied. They are typically reproduced
 by a combination of line-of-sight (LOS) velocity and magnetic field
 gradients \citep[see e.g.,][]{Illing:1975,
Sanchez-Almeida:1988ab,Grossmann-Doerth:1989aa,Solanki:1989aa,Rueedi:1992aa,Lopez-Ariste:2002aa}.

On the other hand, there is increasing observational evidence
showing anomalous circular polarization signals that are far from
having two antisymmetric lobes. 
Such nonstandard profiles can have several shapes: only one lobe
\citep[][]{Sanchez-Almeida:1996, Sigwarth:1999,Grossmann-Doerth:2000aa,Sainz-Dalda:2012aa}; 
two prominent lobes of the same
sign \citep[hereafter double-peak
profiles, see e.g.,  ][]{Grossmann-Doerth:2000aa,Sigwarth:2001aa,Lopez-Ariste:2009aa};
three (or four) lobes of
alternate signs, which would be called type III (or IV) following
\cite{Sanchez-Almeida:1996} classification, or mixed-polarity types following \cite{Sigwarth:2001aa};
or even three lobes with {nonalternate} signs 
\citep[][]{Sanchez-Almeida:2008,Kiess:2018aa}. 

Some of the abnormal Stokes V profiles may be emulated 
 by a multi-component model atmosphere combining
 quasi-ordinary Stokes V profiles with different amplitudes and
spectral shifts, while others 
have symmetry properties challenging such approach.
 In any case, one has to consider that
almost any polarization profile can be explained by more than one solar
 atmospheric configuration (spectral ambiguity). Hence, mimicking
 profiles cannot necessarily be considered as proof that a given model is accurate. As highlighted in
 \cite{Carlin:2017aa}, the spectral ambiguity is a
 problem when investigating scattering polarization signals because 
they can be affected by many factors (particularly by atomic polarization, Hanle
 effect, and chromospheric velocity gradients). If definitive proof of accurate modeling is to
 be obtained, we need to define discrimination techniques that allow us to clarify, and
 if possible disentangle, the physical mechanisms in action
 \citep{Carlin:2015aa}. A posteriori, such constraints could be
 explicitly introduced in inversion codes to make them immune to
  spectral ambiguities.    

In this regard, one possibility is to study anomalous circular
polarization (CP) signals
 to bypass the
complex angular behavior of radiation field anisotropy modifying
linear polarization (LP), 
to complement the information in LP,
and to avoid ambiguities. 
With this idea in mind, we are analyzing, in a separate publication,
 quiet-sun observations of Stokes V signals taken with the ZIMPOL
 spectropolarimeter \citep{Ramelli:2010aa} at GREGOR \citep{Volkmer:2007}
in chromospheric spectral lines forming in nonlocal
 thermodynamical equilibrium (NLTE). We find that they 
exhibit continuous spatial 
 transformations between near-ubiquitous
 ``Q-like'' profiles (three lobes with alternate signs) and
  double-peak profiles. 
In that process, anomalous LP signals and chromospheric
  velocity gradients were identified and need to be explained. 

\cite{Grossmann-Doerth:2000aa} and
\cite{steiner2000} studied the anomalous Stokes V signals in LTE 
considering a model of a static magnetic canopy
above a nonmagnetic moving layer and emphasizing the
presence of their abrupt interface, the magnetopause. This
canopy model seeks to have a nonmagnetic moving layer spatially separated from
a magnetic layer \citep{Grossmann-Doerth:1989aa}, thus assuring the presence of 
simultaneous gradients of magnetic and velocity fields along the
LOS. The need for this combination to explain the observed
 Stokes V asymmetries and net
circular polarization (NCP) was proposed by
\cite{Illing:1975} and supported by the work of 
\cite{Sanchez-Almeida:1989} under the assumptions of 
relatively small velocity gradients. However, \cite{Lopez-Ariste:2002aa} showed that in a
general case with arbitrary velocity gradient its role in
generating NCP is key and more fundamental than that of the
magnetic field. The association between velocity gradients and the
formation of solar CP has been well studied
in different physical conditions by several authors \citep[e.g., ][]{Martinez-Pillet:1994aa, del-Toro-Iniesta:2001aa, Borrero:2010aa}.

Two limitations of the works cited above is that they try to explain the formation of
 Stokes V profiles assuming LTE and without considering the role of atomic polarization. 
This is not suitable for our purposes because spectral lines with 
anomalous polarization profiles may form in NLTE, typically in the upper
photosphere or above (e.g., the Sodium D lines). 
The action of the radiation field in the atomic system
(i.e., the generation of atomic polarization)
requires in these cases a NLTE solution to the statistical equilibrium
equations (SEE).
Though NLTE effects may be ruled out for
spectral lines forming in volumes that are sufficiently dense so as to be collisionally depolarized, 
the boundary between the two regimes is unclear, 
strongly line-dependent, and probably also spatio-temporally dependent.
Assuming LTE leads to a second inconsistency: previous works disregard
atomic orientation while assuming a radiation field modified by 
velocity gradients (i.e., with NCP). This has the contradictory effect of
inducing atomic orientation, because the Doppler-induced spectral
imbalance of intensity modulates the atomic density matrix by means of 
the radiation field tensor in the SEE 
\citep[see Eq. (7) in][]{Bommier:1997ab}.
Hence, if significant velocity gradients are invoked to explain the
circular polarization, the role of atomic orientation should also be
investigated. 


In the present paper our goal has
  been to develop a suitable model accounting for the observed circular
  polarization profiles, especially the relatively frequent Q-like
  profiles, without neglecting the physics of NLTE and atomic polarization.
In Section 2, we introduce
the minimal arguments and
physical ingredients to explain 
Stokes V signals with atomic orientation. In Section 3, we present
 a simple radiative transfer model and propose a first spectral mechanism able to 
 produce double-peak V profiles in nonmagnetic atmospheres.
In Section 4, we use these latter concepts to study with detailed calculations a second
 spectral mechanism that explains Q-like Stokes V signals without
 Zeeman splitting. Sections 5 and 6 demonstrate the ability and convenience of our
  theoretical description in explaining any CP
  signal.
\begin{figure}[!h]
\centering
\includegraphics[scale=1.2]{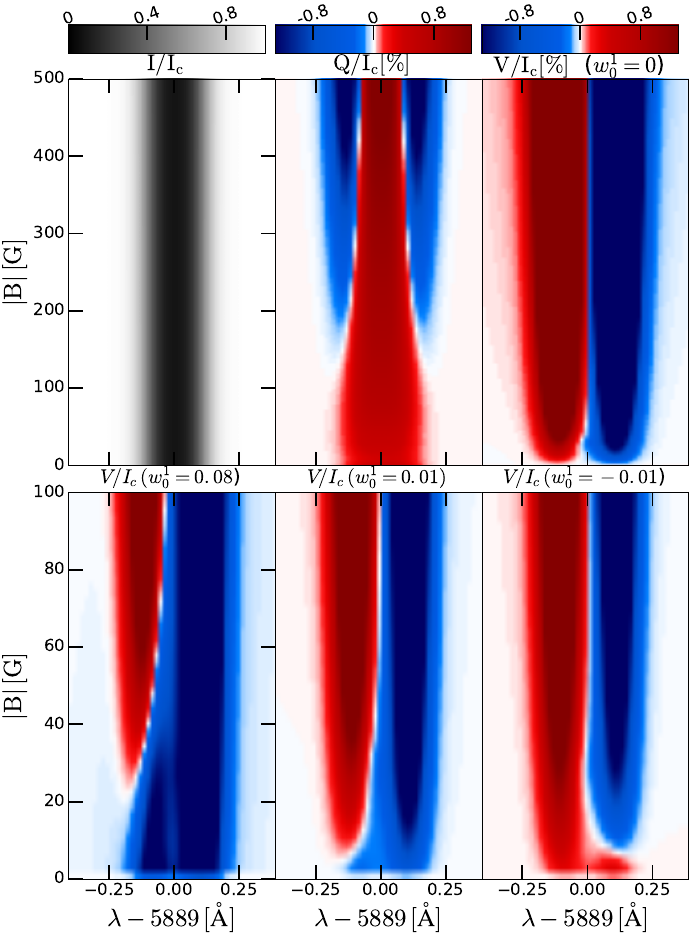}  
\caption{Behavior of $V/I_c$ when varying the magnetic strength and the size and sign of the
  radiation field orientation without gradients along the LOS. All
  polarization panels share the same color scale saturated to $1\%$. Stokes Q and I are plotted as a reference. The example corresponds to the Na {\sc i} D$_2$ line
  modeled in a multilevel atom with hyperfine structure for a magnetic vector inclined $70^o$ from the LOS.}
\label{fig:fig0}
\end{figure}
\section{Preliminars}\label{sec:expectedV} 
\subsection{Atomic orientation}
In a statistical collectivity of scattering atoms, 
an atomic level $J\neq 0$ (or $F \neq 0$ for atoms with
nuclear spin) is oriented when there is
an imbalance of electronic population between Zeeman
sublevels whose magnetic quantum numbers $M$ have opposite signs. This is expressed
as $\rho_{-M-M}(J) \neq \rho_{MM}(J)$ in the
standard representation of the atomic density matrix. In the multipolar representation,
atomic orientation is instead quantified by tensor components of the atomic
density matrix $\rho^K_{Q}(J)$ with rank $K=1$ and $Q=0$ \citep[see Sect.
3.6 of][for details]{LL04}. The SEE indicate that 
atomic orientation can be created by an oriented radiation field,
that is, radiation with net angular momentum, characterized by a
multipolar tensor component $\bar{J}^1_0 \neq 0$. 
For clarity, the generation of atomic orientation is treated in a
separated paper, while here the amount of radiation field orientation is conveniently
parameterized by $w_1=\bar{J}^{1}_0/\bar{J}^{0}_0$, where the mean values of the radiation field
tensor are\footnote{Here, $I_j(j=0,1,2,3)$ are the Stokes parameters
illuminating the scatterer; $\mathcal{T}$ are the spherical tensors for
  polarimetry \citep{Landi-DeglInnocenti:1983};
  $\phi(\nu)$ are the absorption profiles at frequency $\nu,$ and $\vec{\Omega}$
  is the direction of each pumping ray at the scattering point.}:
\begin{subequations}\label{eq:j10j00}
\begin{align}
\bar{J}^{1}_0=& \int d\nu \phi(\nu) \oint
\frac{d\vec{\Omega}}{4\pi}\sum^{3}_{i=1}\mathcal{T}^{1}_0(i,\vec{\Omega})
{\mathbf I}_i(\vec{\Omega},\nu) ,\\
\bar{J}^{0}_0=& \int d\nu \phi(\nu) \oint
\frac{d\vec{\Omega}}{4\pi} {\mathbf I}_0(\vec{\Omega},\nu).
\end{align}
\end{subequations}
These equations show that atomic orientation is induced by
spectral asymmetries, particularly in intensity and Stokes V. 
This explains why NCP is one of the proxies to radiation
field orientation. NCP can be defined as\footnote{
Another definition is \citep[][]{Landolfi:1996aa}:
\begin{align}\label{eq:ncp}
\rm{NCP} &=  \int d\lambda \,V(\lambda) / \int d\lambda \left[ I_c -I(\lambda) \right],
\end{align}
with $I_c$ being the continuum intensity. We do not use this
definition because we think that the NCP should be normalized to the total
number of photons, not to the depression of the spectral line.}:
\begin{align}\label{eq:ncp_ed}
\rm{NCP} &=  \int d\lambda \,V(\lambda) / \int d\lambda I(\lambda),
\end{align}
or simply as the numerator when comparing two V profiles whose total
number of intensity photons are similar.

\subsection{Expected vs. observed weak-field V/I signals}\label{sec:ori2peaks}

Double-peak profiles and Q-like profiles are the paradigm for nonstandard
CP because, having the possibility of being symmetric, they strongly differ
  from the antisymmetric Zeeman-like signals. 
Hence, when trying to explain the anomalies,
one of the first logical doubts is whether the Zeeman effect
is key. We know that in weak fields the transversal Zeeman
effect is of second order and as a consequence the typical
three-lobe LP signatures do not normally appear for fields
below $100$ G. Indeed, in Na{\sc i} D$_2$ the lateral lobes in Stokes Q
(formed by the Zeeman $\sigma$ components) are comparable to the central $\pi$
component only for field strengths above $200$ G (Fig. \ref{fig:fig0},  upper central
panel). On the contrary, the longitudinal Zeeman effect
is already measurable for strengths as small as $10$ Gauss (Fig. \ref{fig:fig0}, right upper
panel). Hence, the Zeeman effect seems necessary to
explain double-peak patterns. However, the conservation of angular momentum
requires any pair of $\sigma_+,\sigma_{-}$ components to  always have opposite sign.
 Consequently, something else than just the Zeeman effect is required to produce
such anomalous profiles. 

\begin{figure}[!h]
\centering
\includegraphics[scale=0.85]{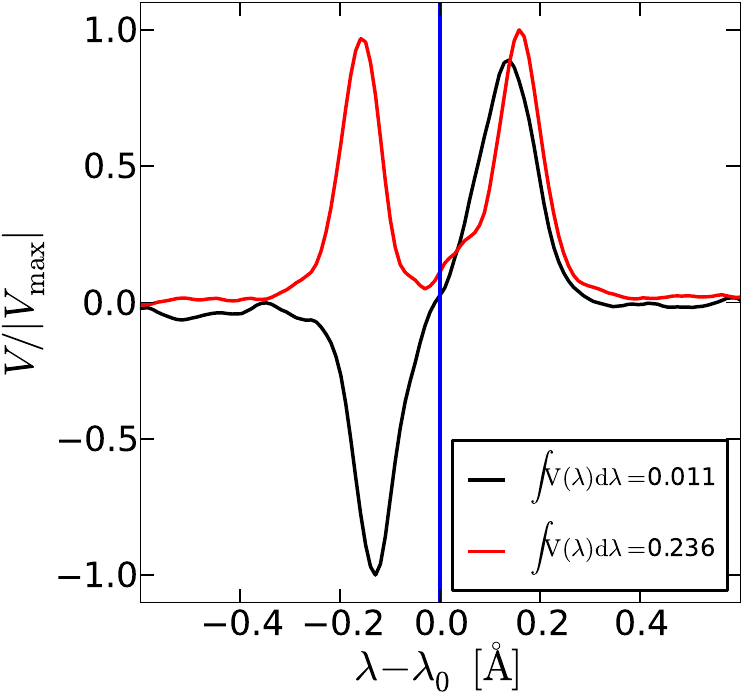}   
\caption{Observed double-peak Stokes V profile (red) vs. standard V
  profile (black) normalized to their maximum amplitudes ($0.2 \%$
  and $0.1\%$, respectively) in the Na {\sc i} D$_1$ line. The
  profiles were measured
 at the THEMIS solar
telescope. See similar profiles (taken in same campaign as ours) in Fig. 5 of
\cite{Lopez-Ariste:2009aa}. We also found them in Fig. 3 of \cite{Bommier:2002aa}.}
\label{fig:fig0b}
\end{figure}

On the other hand,  the NCP of nonantisymmetric V signals can induce
orientation, as mentioned above. 
Indeed, the mere existence of
V signals with only one sign should imply a significant increment of
the atomic orientation because they can carry an order of magnitude more NCP than normal
signals. To see this, one can calculate the integral under the curve of a V
double-peak profile and compare it with the same integral for a
realistic standard antisymmetric
profile of the same amplitude. As shown by Fig. \ref{fig:fig0b}, the
former integral is $21$ times larger than the latter, simply because 
the amount of signal that was subtracted
in the integral of the antisymmetric profile is now contributing positively. In the
particular case of the observation in Fig. \ref{fig:fig0b}, the NCP 
(which furthermore accounts for differences in
intensity and polarization amplitudes) is $22$ times larger in the
double-peak profile.
This supports the idea that the existence of atomic
orientation in the scatterers is an ingredient 
to be considered in the formation of anomalous V profiles.

However, atomic 
orientation cannot act as a spectral mechanism inverting
 only one of the peaks of Stokes V unless 
such an effect is replicated in the optical coefficients of the radiative
transfer equation (RTE). This is not
possible because the increment of atomic orientation always enhances the
contribution of one set of Zeeman sigma components (e.g., the $\sigma_{+}$) at the expense of the
other. For weak fields, the spectral result is the fusion of both peaks in a single asymmetric peak\footnote{Note that the asymmetry of the
 resulting coefficient depends on the level of orientation, but also on the
 Zeeman splitting. In an extreme case in which orientation is so
 strong that depletes
 one of the sigma components, the result would be a symmetric peak
 shifting away from line center as the field increases, hence resulting in a
completely asymmetric signal in strong field if the axis to measure symmetry
is line center.}. This is illustrated by a comparison between the right upper
panel and the lower panels of Fig. \ref{fig:fig0}. Here we have
induced atomic orientation in the atoms of a slab model by pumping it with an oriented radiation field.
Thus, the ordinary signals are such that, when the Zeeman
splitting tends to zero, the sigma components
overlap in frequency, either canceling each other in the absence of
orientation (see upper right panel of Fig. \ref{fig:fig0} for $B=0$),
or resulting in a single (though irregular/asymmetric) lobe whose absolute amplitude is 
proportional to the amount of orientation (see lower panels of
Fig. \ref{fig:fig0} for $B\in[0,10]$ G).
 The difference between the calculations in the left and central lower
 panels is the amount of orientation: the comparison shows that larger
orientation produces visible morphological effects persisting in larger field strengths.

In summary, the larger the magnetic field, the larger the Zeeman
splitting, and the less effective the cancelation of sigma components;
consequently, more atomic orientation is required to induce the same
level of asymmetry in the corresponding Stokes V profile. 
The other side of this sort of ``law of scalability'' or
self-similarity involves the
causality existing between the increment of the velocity gradients (or Doppler shifts) and the
amount of orientation that they induce: the larger the former, the
greater the latter. Thus, the standard expectation is that
orientation wins the morphological battle in layers where macroscopic motions are
 large, while magnetic fields
 and depolarizing collisions are weak (e.g., in the chromosphere). 
At the end of the paper, the role of atomic
orientation in intermediate and strong fields is reconsidered in light of our results.

For now, we only conclude that despite the fact that  orientation effectively modifies the
spectral profile and can induce 
 asymmetries in relation to the velocity
gradient, it alone cannot
invert the sign of only one peak of the V signals. Hence, apart from Zeeman splitting and atomic orientation, we still need a third physical ingredient, a spectral mechanism. In the following
we study two physical setups able to emulate the
observed V signals. The  spectral mechanism of the first one is dichroism
induced by atomic orientation. The second one also
requires dichroism, but this time combined with uneven excitation of the
Zeeman components and, though it can be mimicked without atomic
polarization, the latter modifies the signal significantly.

\section{Radiative transfer model}\label{sec:equations} 
We start posing the RTE for the Stokes vector
$\mathbf{I}=(I,Q,U,V)^T$ along a ray $\vec{\Omega}$ as a function of
the optical depth $\tau_{\nu} $:
\begin{equation}
\centering
  \frac{\rm d \mathbf I}{{\rm d} \tau_{\nu}}  = \frac{\mathbf K
    \mathbf I-\boldsymbol{\epsilon}}{\eta_I}   \,,
\label{eq:RTE}
\end{equation}
where all the quantities depend on $\tau_{\nu} $ and frequency $\nu$, giving $d \tau_{\nu}=-\eta_I ds$ with
$s$ the geometrical distance along the ray. The emission vector $\boldsymbol{\epsilon}=(\epsilon_I,\epsilon_Q,\epsilon_U,\epsilon_V)^{T}$
quantifies the total generation of intensity and polarization in each plasma element, while
the propagation matrix is
\begin{equation*}
  \mathbf K = \begin{pmatrix}
      \eta_I &  \eta_Q &  \eta_U & \eta_V  \\
      \eta_Q &  \eta_I &  \rho_V & -\rho_U \\
      \eta_U & -\rho_V &  \eta_I & \rho_Q  \\
      \eta_V &  \rho_U & -\rho_Q & \eta_I 
               \end{pmatrix},
\label{eq:matrix_K}
\end{equation*}
with $\rho$ being the anomalous dispersion terms, $\eta_I$ the total
absorption (line plus continuum), and $\eta_{Q,U,V}$ the dichroism coefficients. 

A detailed explanation of the anomalous Stokes V signal can be
achieved ignoring the negligible contributions of stimulated emission and second-order
terms (of the kind $\eta_i I_k$ and $\rho_i I_k$ when both $i,k\neq 0$ ) in the RTE. This leads to
\begin{subequations}\label{eq:simple_rte}
\begin{align}
& \frac{\rm d I}{{\rm d} \tau_{\nu}}  = I - S ,\\
& \frac{\rm d V}{{\rm d} \tau_{\nu}}  = V - \frac{\epsilon_V-\eta_VI}{\eta_I,} 
\end{align}
\end{subequations}
where $S=\epsilon_I/\eta_I$ is the total intensity source function. To solve the equations,
we  consider an arbitrary portion of atmosphere with geometrical size $\Delta
s$ and finite optical
thickness $T_\nu=\Delta s \cdot\eta_I$ (the frequency subindex 
indicates that each wavelength of the radiation perceives a different
opacity). The
solutions to the RTE are then
\begin{subequations}\label{eq:sol_rte1}
\begin{align}
& I(\tau_\nu)= I_0 \, e^{-(T_{\nu}-\tau_{\nu})} +\int^{T_{\nu}}_{\tau_\nu}\rm d \tau'_{\nu} \,S\,e^{-(\tau'_{\nu}-\tau_{\nu})} ,\\
& V(\tau_\nu)= V_0 \, e^{-(T_{\nu}-\tau_{\nu})} +\int^{T_{\nu}}_{\tau_\nu}\rm d
\tau'_{\nu} \left[\frac{\epsilon_V 
-\eta_V I(\tau'_{\nu})}{\eta_I}\right]  e^{-(\tau'_{\nu}-\tau_{\nu})},
\end{align}
\end{subequations}
with $(I_0,V_0)$ being the boundary illumination entering the
atmosphere\footnote{We model the formation region between vacuum and a maximum optical depth after which an $I_0$ is present.} along the
LOS. The outgoing radiation traveling towards the
observer is then
\begin{subequations}\label{eq:sol_rte2}
\begin{align}
& I(0)= I_0 \, e^{-T_{\nu}} +\int^{T_{\nu}}_{0}\rm d \tau'_{\nu} \,S\,e^{-\tau'_{\nu}} ,\\
& V(0)= V_0 \, e^{-T_{\nu}} +\int^{T_{\nu}}_{0}\rm d
\tau'_{\nu} \left[\frac{\epsilon_V 
-\eta_V I(\tau'_{\nu})}{\eta_I}\right]  e^{-\tau'_{\nu}}.
\end{align}
\end{subequations}
Introducing Eq.(\ref{eq:sol_rte1}a) into Eq.(\ref{eq:sol_rte2}b) and
assuming that all the quantities but the source function are constant
with position inside the slab we obtain
\begin{equation}\label{eq:sol_rte3}
\begin{aligned}
V(0) &= V_0 \, e^{-T_{\nu}} + \frac{(1-e^{-T_{\nu}})\epsilon_V 
-\eta_V I_0 T_{\nu} e^{-T_{\nu}}}{\eta_I} -\\
&-\frac{\eta_V }{\eta_I}\int^{T_{\nu}}_0\rm d\tau'_{\nu} \int^{T_{\nu}}_{\tau'_\nu}
\rm d \tau''_{\nu} S(\tau''_{\nu}) e^{-\tau''_{\nu}}
\end{aligned}
.\end{equation}
In general, the quantities $I_0=I_0(\nu)$ and $S=S(\nu,\tau_{\nu})$ have spectral 
structure. For consistency with the previous step
we assume that $S$ is constant with depth. Working with the total
optical coefficients $\bar{\eta_i}=\Delta s \cdot \eta_i$ and
$\bar{\epsilon_i}=\Delta s \cdot\epsilon_i$, 
 integrating, and reorganizing, we arrive at the useful expression,
\begin{equation}\label{eq:sol_rte5}
 V(0)= V_0 \, f_0 + \left(\beta_1 f_1 - \beta_0 f_0\right) 
,\end{equation}
where 
\begin{subequations}\label{eq:fs}
\begin{align}
 \beta_1=& (S_V-S) \bar{\eta}_V \quad ; \quad  f_1  (T_{\nu})=
 \frac{1-e^{-T_{\nu}}}{T_{\nu}} ,\\
 \beta_0=& (I_0-S) \bar{\eta}_V \quad ; \quad   f_0 (T_{\nu})=  e^{-T_{\nu}},
\end{align}
\end{subequations}
and $S_V=\bar{\epsilon}_V/\bar{\eta}_V$.
Finally, the emergent fractional polarization is\begin{equation}\label{eq:VoverI}
\frac{V(0)}{I(0)} =\frac{V_0 \, f_0 + \left(\beta_1 f_1 -
     \beta_0 f_0\right)}{I_0 \, f_0 + S(1-f_0)}
.\end{equation}
This model is also valid for LP substituting
the letter V by Q or U in the corresponding quantities.
Equation (\ref{eq:sol_rte5}) has three contributions: 
a first one, due to the boundary illumination, becomes quickly negligible for
 $T_{\nu}\gtrsim 2$; a second one depends
on the relation between $S_V$ and $S$
(contained in $\beta_1$);
and the last one depends on the relation between $I_0$ and $S$
(contained in $\beta_0$). 
Each term includes a transfer function $f$ depending on $T_{\nu}$. 
The dependence on magnetic field and atomic orientation is contained in the
optical coefficients.

\section{Spectral mechanism without Zeeman splitting}\label{sec:spectral_1} 
\begin{figure}[!t]
\centering
\includegraphics[scale=0.5]{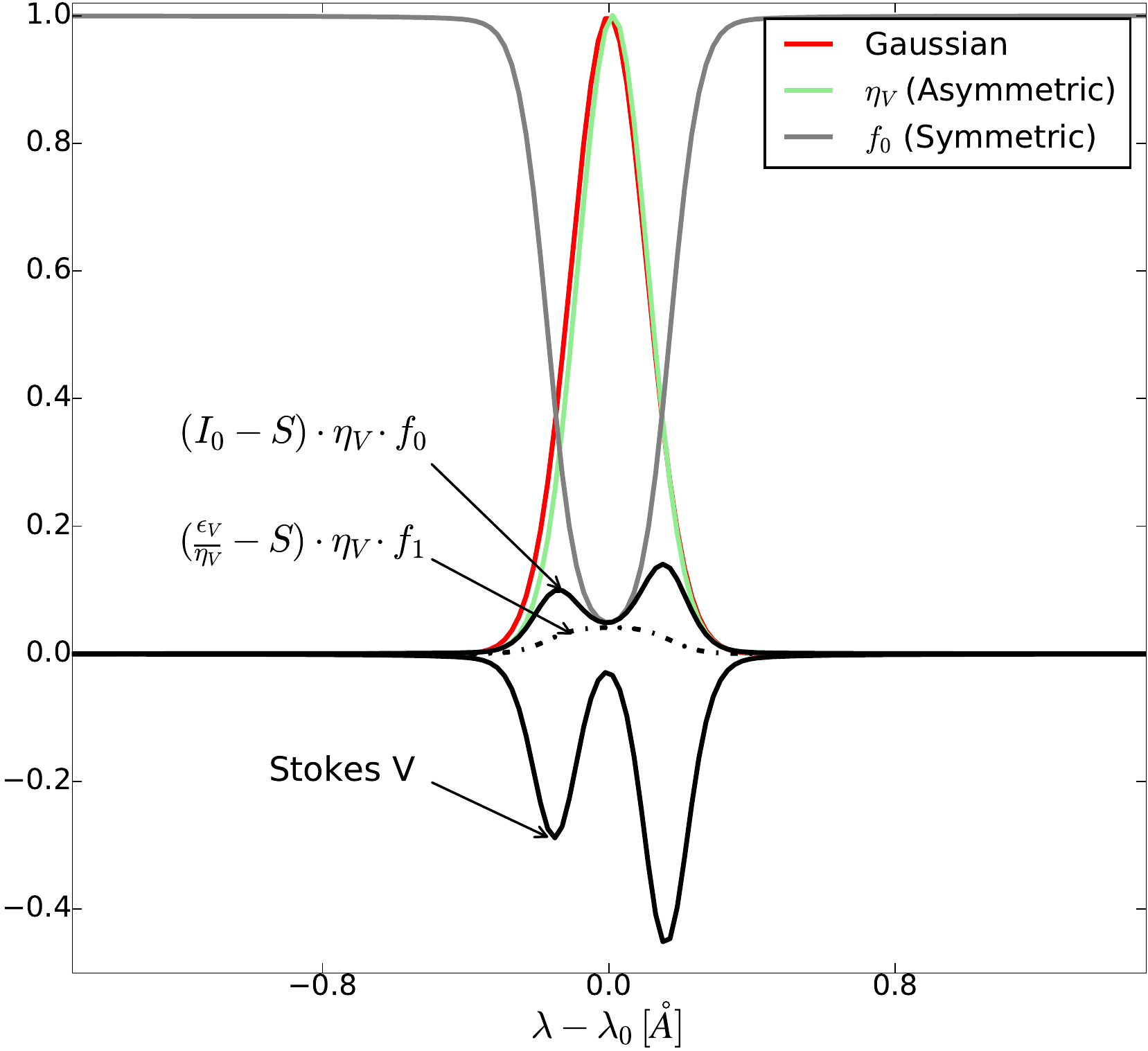}   
\caption{Generation of an asymmetric double-peak profile with $B=0$
  and atomic orientation ($w_1=5\%$) following
 Eq. (\ref{eq:sol_rte5}). The quantities represented in the
 vertical axis are labeled inside the figure but only their shapes are important here. The green curve is
 normalized to one, and the black curves are multiplied by the same constant, in order to
 plot everything together. The factors 
$\epsilon_V$ and $\eta_V$ were calculated for the Na {\sc i} D$_2$ line in a multilevel atom
with HFS. }
\label{fig:fig1}
\end{figure}

We now consider an academic situation in which: a) the magnetic field
is zero, which implies optical coefficients with a
single lobe, as explained in Sect. \ref{sec:expectedV}; b) a given
amount of atomic orientation has been induced by pumping with a
radiation field $J^1_0\neq0$; and c) the illumination along the LOS is
spectrally flat. In this situation Equation (\ref{eq:sol_rte5})
can emulate all the spectral features
of double-peak and Q-like Stokes V signals, namely: the double peak structure, the asymmetry
and distance between the peaks, and the depth of the central dip. 
Figure \ref{fig:fig1} illustrates how such a spectral mechanism
 works. The factor $f_0$ is equal to one in
the wings and approaches zero in the core, such that when multiplied by
$\bar{\eta}_V(\nu)$ can produce a
double-peak profile.
\begin{figure}[!t]
\centering
\includegraphics[scale=0.55]{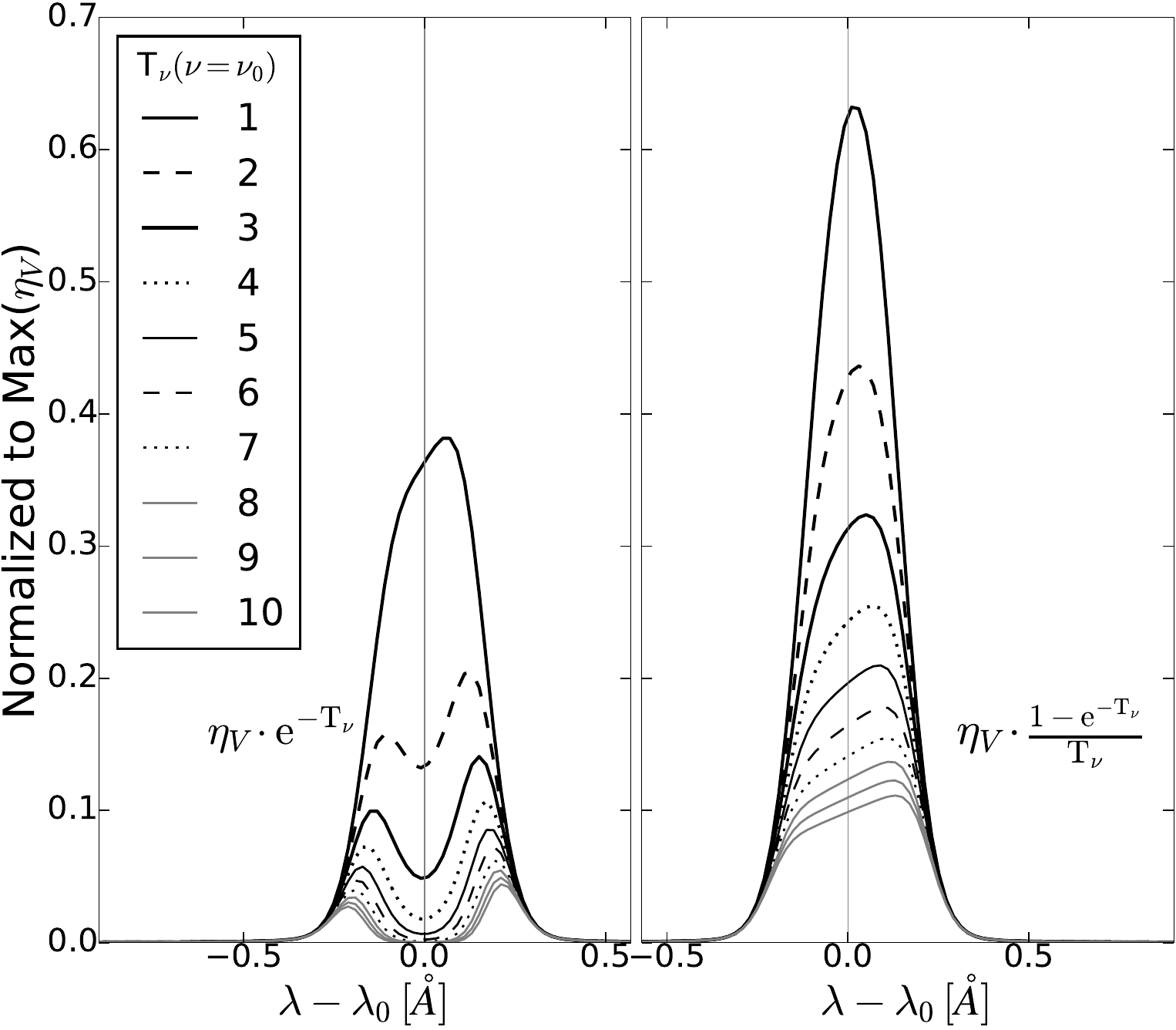}   
\caption{Origin of the double-peak spectrum in Fig. \ref{fig:fig1} as a function of the
  optical depth of the slab $T_{\nu}$, following Eq. (\ref{eq:sol_rte5}). Left: Factor generating a
  double-peak profile. Right: Factor opposing the generation of the peaks.}
\label{fig:fig2}
\end{figure}
\begin{figure*}[t!]
\centering$
\begin{array}{cc} 
\includegraphics[scale=0.5]{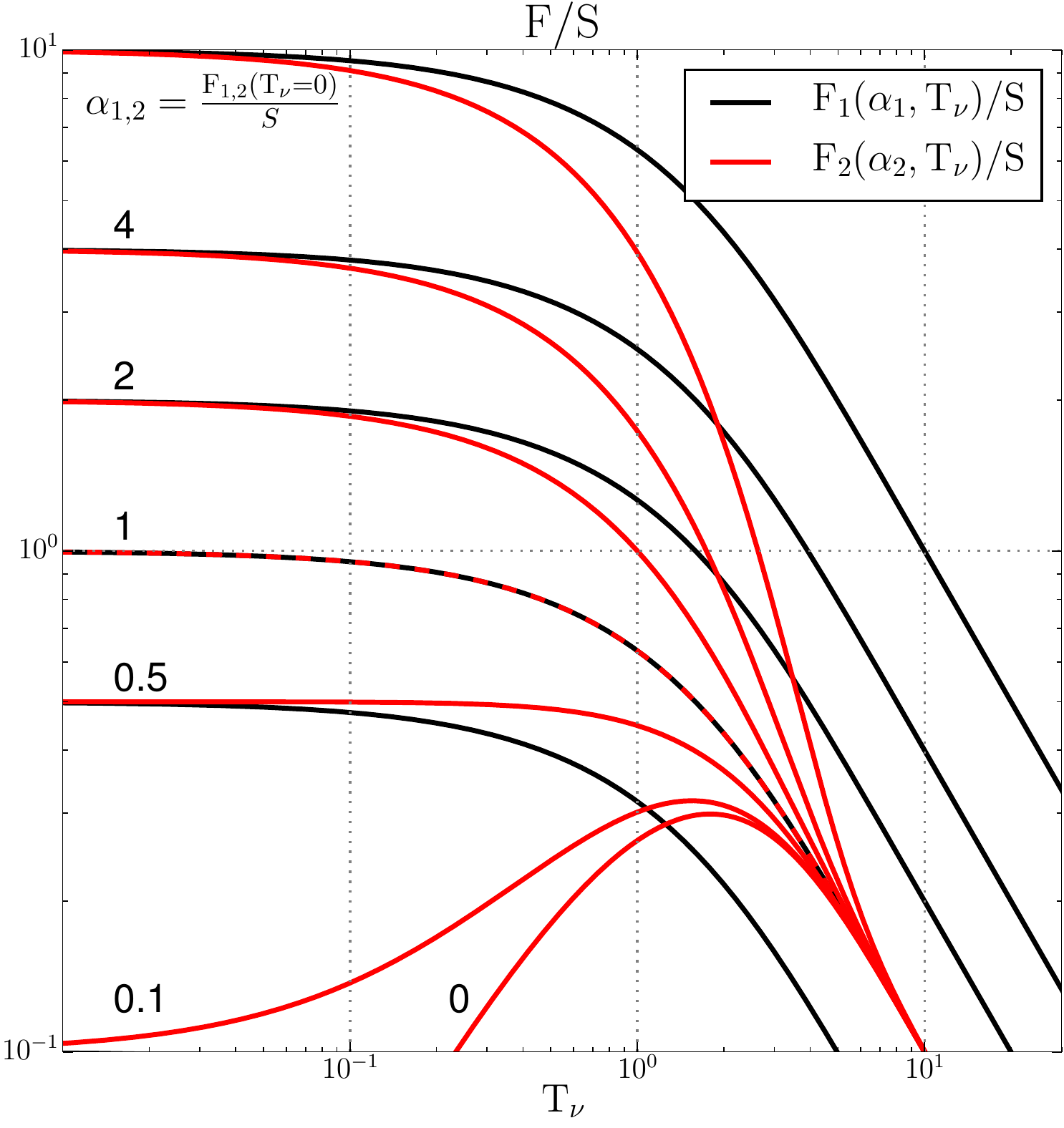}   & 
\includegraphics[scale=0.5]{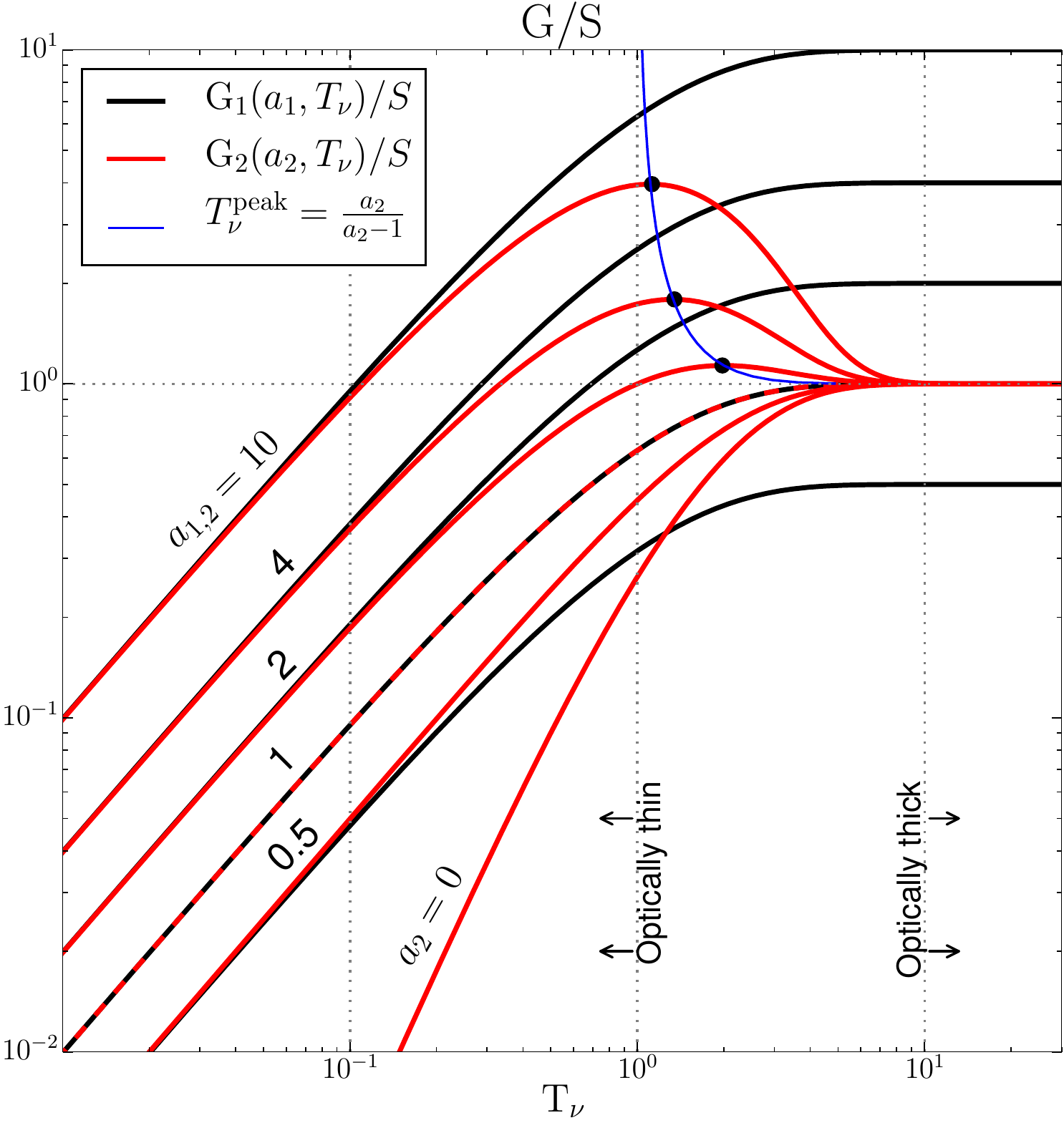}\\ 
\end{array}$
\caption{Left: Variation of $F_1/S$ and $F_2/S$ with optical depth for different
   levels of V source function ($\alpha_1=S_V/S$) and intensity
   ($\alpha_2=I_0/S$), respectively. Right: Same as left panel
   but for $G_1/S$ and $G_2/S$ with their corresponding parameters
   $a_1$ and $a_2$. Limiting values are:
   $F/S(\alpha_1,T_{\nu}\geq 10)=\alpha_1/T_{\nu}$,
   $G/S(a_1,T_{\nu}\geq 10)=\alpha_1$ and  $G/S(a_{1,2},T_{\nu}<< 1)=a_{1,2}T_{\nu}.$}
\label{fig:fig3}
\end{figure*}
Subtracting the result from the
 term depending on $f_1$ amplifies the contrast (and the eventual
 asymmetry) between the peaks and produces Stokes V. 
The term $f_0$ alone only generates two peaks when the maximum total opacity
of the atmosphere is above one, as shown in the left panel of Figure
\ref{fig:fig2}. On the contrary, the factor $f_1$ alone can only have one
peak (see left panel of Figure
\ref{fig:fig2}) and its role is to increase Stokes V in the line center,
thus modulating the central dip and the visibility (contrast) of the
double peaks. What we are describing here is nothing but a balance or competition
between emissivity (included in the factor $\beta_1f_1$) and dichroism
(dominating the factor $\beta_0 f_0$). The signal with two peaks of
the same sign is possible because the transfer function $f_0$ 
shapes the dichroic contribution in its central part, which is observable in
Stokes V when emissivity is low enough. Due to this sort of
self-absorption in polarization, the
explanation of the V peaks in this situation emphasizes the role of dichroism.

In a static unmagnetized media such as the one considered, 
the asymmetry between the V peaks is due to the optical
coefficients. The deviation that these coefficients can have from a
Gaussian comes from atomic orientation and from the asymmetric spectral
distribution of Zeeman components (due to HFS in case of our
archetypal Sodium atom).
In the numerical experiment of Fig. \ref{fig:fig1}, only $\eta_V$ is asymmetric, while $\epsilon_V$
remains practically symmetric. The particular asymmetry of one
or other coefficient can change depending on the atomic
species, the transition, and the solution to the SEE. Figure \ref{fig:fig1} shows that the deviation that $\eta_V$
shows from the symmetric (Gaussian) profile is transformed efficienly into a significant 
asymmetry in Stokes V during the radiative transfer. This happens at the level of the atomic
collectivity, without the need for macroscopic motions.

\subsection{The essential behavior of dichroism}
The simplicity of our model allows us to study analytically the action of dichroism and its
balance with emissivity.
In our context, the only ways of having dichroism, that is, a selective
 absorption of polarization states in an atom, are to have a Zeeman
 splitting making the absorption selective in wavelength, and/or
 to have atomic polarization (the only possibility in
 unmagnetized fluids or in magnetic plasmas when the magnetic field
 is not longitudinal).
In order to decipher the conditions in which the paradigmatic double-peak signals 
are formed with our first (non-magnetic) spectral
mechanism, we reformulate Eq.(\ref{eq:sol_rte5}) using
  quantities normalized to the source function:
\begin{equation}\label{eq:new2}
V(0)= V_0 \, e^{-T_{\nu}} + \left( F_1 -
    F_2 \right) \cdot \bar{\eta}_V  \\
,\end{equation}
where 
\begin{subequations}\label{eq:fs2}
\begin{align}
 \frac{F_1(\alpha_1,T_{\nu})}{S}&= \alpha_1\cdot\frac{1-e^{-T_{\nu}}}{T_{\nu}} ,\\
\frac{F_2 (\alpha_2,T_{\nu})}{S}&=  \frac{1-e^{-T_{\nu}}\left [ 1+T_{\nu}(1-\alpha_2)\right]}{ T_{\nu} },  
\end{align}
\end{subequations}
\begin{equation}\label{eq:alfa2}
\alpha_1= S_{V}/S \quad ; \quad \alpha_2=I_0/S
.\end{equation}
We now simplify these expressions by approximating the optical 
coefficients using spectrally symmetric Gaussians because there is no
Zeeman splitting, which allows
us to make the following simplifications:\footnote{In this step we are implicitly assuming that the
  ratio between the continuum and the line absorption for intensity is
  negligible (strong line), with the only purpose of getting simpler
  expressions.}
$\bar{\eta}_V
(\nu)=(\eta^{l}_V/\eta^{l}_I)\bar{\eta}_I(\nu)=(\eta^{l}_V/\eta^{l}_I)T_{\nu}$
and similarly $\bar{\epsilon}_V
(\nu)=(\epsilon^{l}_V/\eta^{l}_I)T_{\nu}$, with the index $l$
indicating line coefficients. This simplification is only used in the
following few paragraphs for illustrative and explanatory purposes. 
The new formulation is then
\begin{equation}\label{eq:new3}
V(0)= V_0 \, e^{-T_{\nu}} + \left( G_1 -
    G_2 \right) \cdot \frac{\eta^{l}_V}{\eta^{l}_I}  \\
,\end{equation}

\begin{subequations}\label{eq:fs3}
\begin{align}
\frac{ G_1  (a_1,T_{\nu})}{S}&= a_1\cdot (1-e^{-T_{\nu}}) ,\\
\frac{ G_2 (a_2,T_{\nu})}{S}&=  1-e^{-T_{\nu}}\left [ 1+T_{\nu}(1-a_2)\right],   
\end{align}
\end{subequations}

\begin{equation}\label{eq:alfa3}
a_1= S^l_{V}/S \quad ; \quad a_2=I_0/S
,\end{equation}
where the dependence of the
polarization on frequency and optical depth is now exclusively contained in 
$G_1$ and $G_2$.  The
left panel of Figure \ref{fig:fig3} shows the variation of the two
families of curves $F_1$ (black lines)
and $F_2$ (red lines) with optical
depth for different $\alpha_1$
(relative Stokes V line source function) and $\alpha_2$
(relative incident intensity), respectively. The
right panel shows the same for $G_1$ and $G_2$ with their
corresponding parameters $a_1$ and $a_2$. Subtracting these
latter functions gives Stokes V up to a constant. Thus, small values of $a_1$
shift down\footnote{Note that
 $G_1(a_1,T_{\nu}>10)=a_1$} the black curves in the right panel of
Figure \ref{fig:fig3}, leaving the
emergent polarization dominated by $G_2(a_2,T_{\nu})$.   
Indeed, when the $G_1$ curve goes below the bumps of $G_2$, the 
bumps themselves become the double peaks in the Stokes profile
because each frequency of the light corresponds to a different opacity and
optical depth. 
Namely, fixing the maximum optical depth of the slab in $T_{\nu}$ and varying the frequency
from the core to the wings of the spectral line
 corresponds to sampling the curves in Figure \ref{fig:fig3} toward smaller
optical depths. On the contrary, fixing frequency in the emergent profile and starting to
accumulate optical depth as we penetrate the slab corresponds to
sampling the curves from zero to the maximum optical depth of the slab,
where a boundary $I_0$ illuminates.

 The bumps in $G_2$ are at the same point where the
sensitivity to the illumination is maximum, 
that is, where the depth penetrated into the slab is
around $T_{\nu}=1$. The sensitivity to the illumination quickly saturates around
$T_{\nu}=10$ (clearly, if the slab is too opaque nothing is
transferred). The curves also show that the formation region of the
polarization is essentially between $T_{\nu}=0.1$ and $10$ at every
frequency, as happens with intensity. Since polarized photons
are part of the whole intensity, their formation regions are 
quoted by the same optical depths.

From Eq. (\ref{eq:fs3}b) we calculate the optical depth $T^{\rm{peak}}_{\nu}$
corresponding to the maximum of the
bumps of $G_2$:
\begin{equation}\label{eq:tpeak}
T^{peak}_{\nu}= \frac{a_2}{ a_2-1 },
\end{equation}
which gives the blue curve trending asymptotically to $T_{\nu}=1$ in the right panel of
Fig. \ref{fig:fig3} as $I_0/S$ increases. This simply means that in a real
atmosphere the modifications of the polarization profile due 
to dichroism are maximum around the height of
$\tau=1$ for a given frequency.
 For an illumination varying
with frequency, the same discussion is applied independently to every frequency.
We also note that 
a significant part of the region where $G_2/S$ has negative slope is mapped for optical depths 
$T^{\rm{min}}_{\nu}\approx 3\cdot T^{\rm{peak}}_{\nu}$.

The aim of working with $G_2/S$ ($F_2/S$) has been to expose the dichroic
response and the formation of peaks for relatively strong (weak)
illuminations, as illustrated  by Figure \ref{fig:fig3}. In the right
panel, the bumps in $G_2/S$ 
disappear when the red curves become monotonic ($a_2= I_0/S\leq 1$), that is, when
 the illumination is not able to surpass the emissivity of the
slab. In this (and in any) situation $G_2$ is still not zero. Thus,
provided that there is also emissivity and enough opacity, $G_1$ and $G_2$
can still develop peaks in the Stokes profile when the illumination
drops below a certain limit. The easiest way of seeing this is to
identify such low-intensity peaks with the bumps appearing for
$\alpha_2<0.5$ in $F_2$ (see left
panel of Figure \ref{fig:fig3}). 

\begin{figure}[h!]
\centering$
\begin{array}{c} 
\includegraphics[scale=0.31]{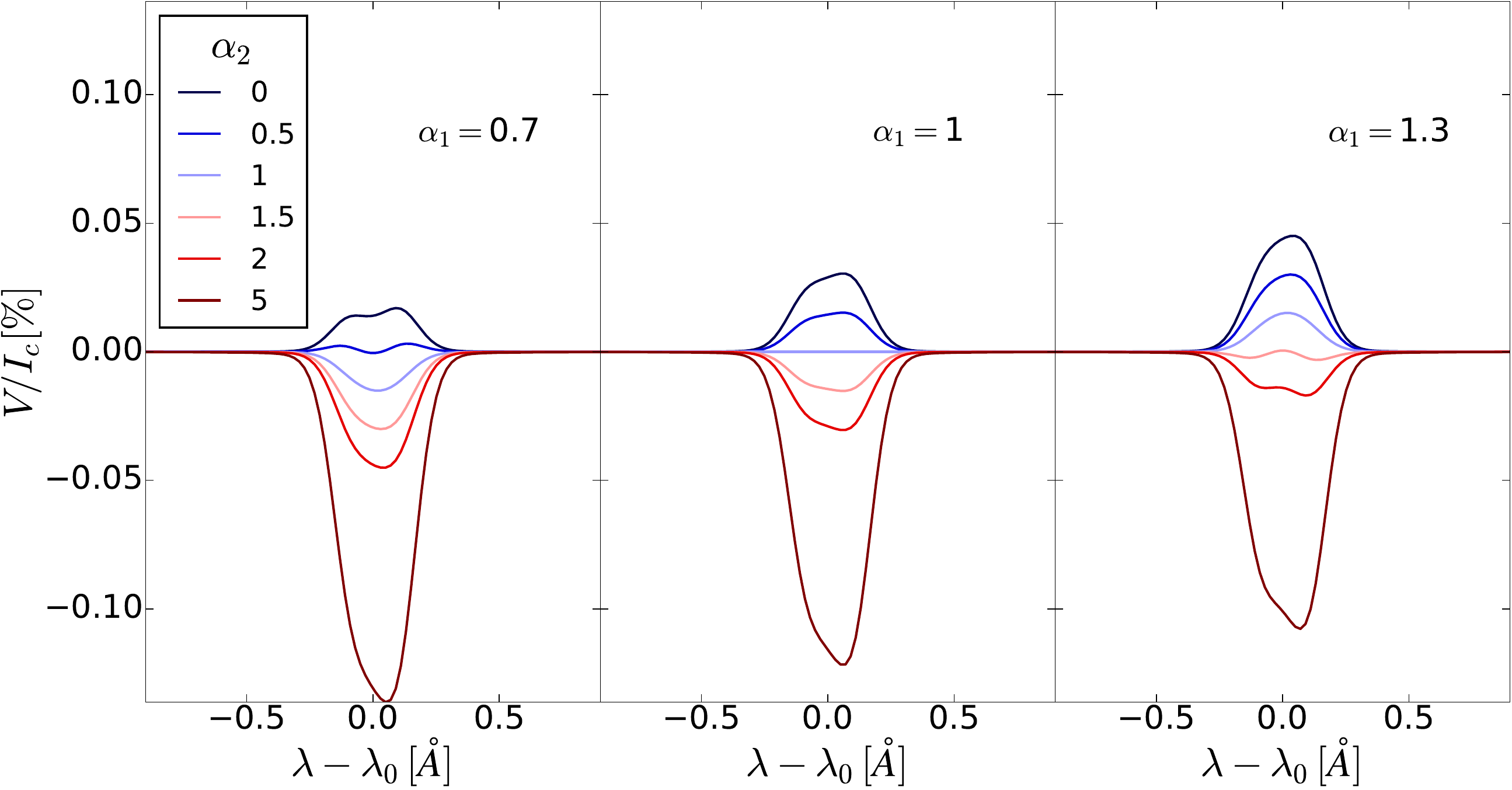}   \\ 
\includegraphics[scale=0.31]{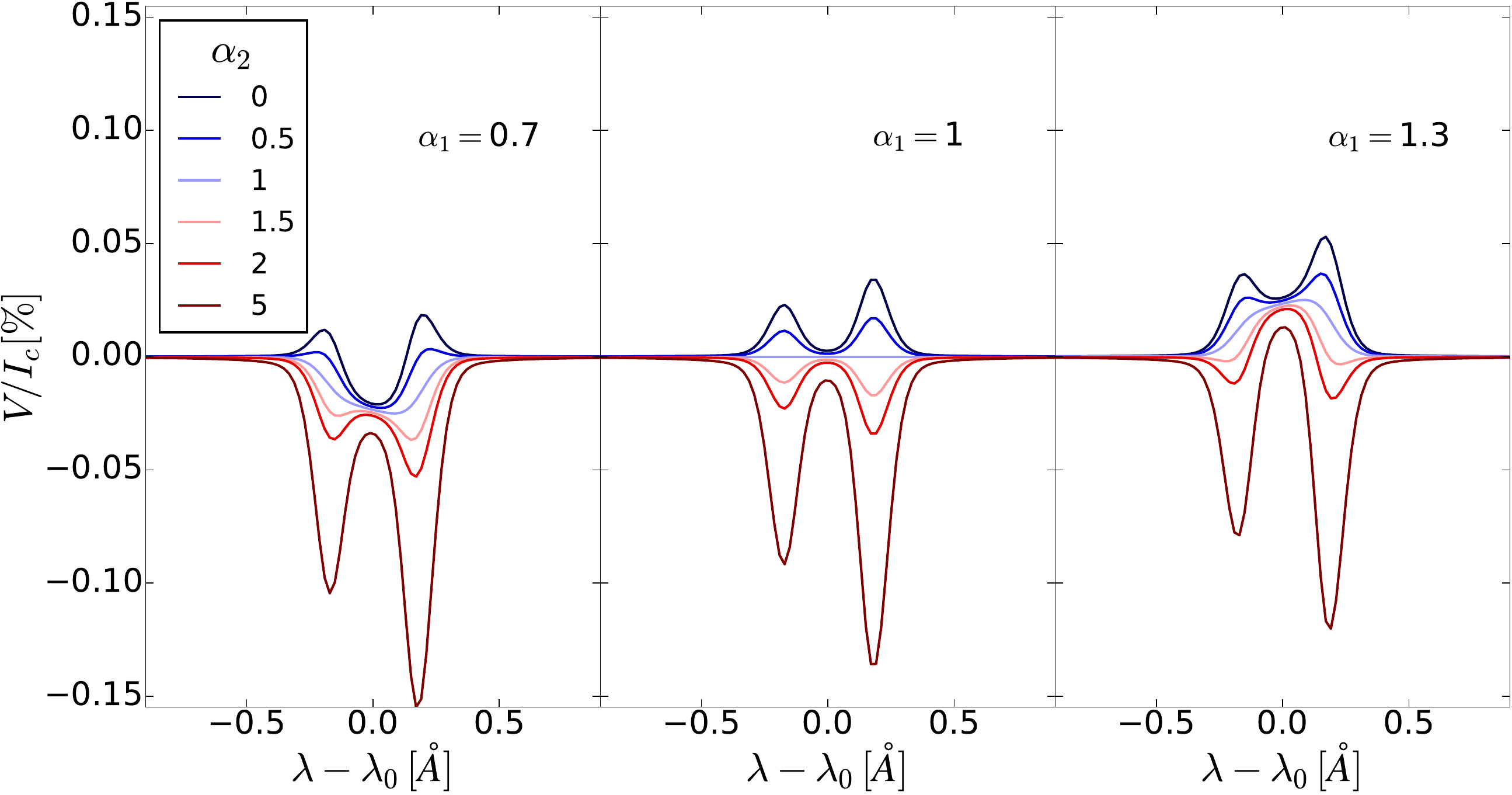} \\
\includegraphics[scale=0.31]{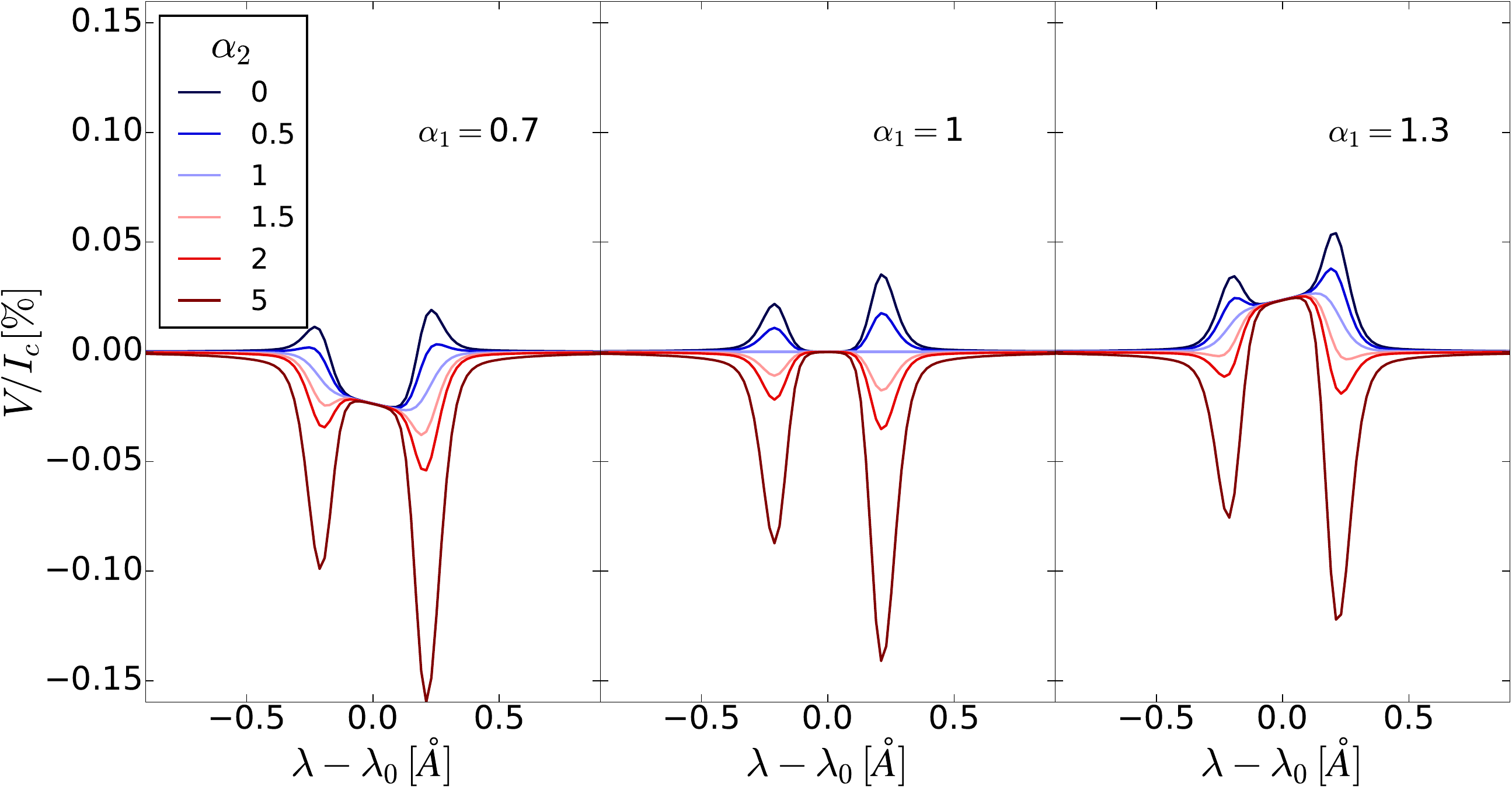}  \\ 
\end{array}$
\caption{Stokes V normalized to the continuum intensity in the Na {\sc i} $D_1$ line for $B=0$, constant
  $w_1$, and for different
  constant boundary illuminations ($\alpha_2=I_0/S$) and positive
  values of $\alpha_1=S_V/S$. Top ($T_{\nu_0}=1$): the opacity is not enough to create the peaks. Middle ($T_{\nu_0}=5$): the
  peaks are clearly formed. Bottom: ($T_{\nu_0}=10$): peaks
  are fully contrasted and start to saturate. The level of the central dip depends
  on the Stokes V emissivity.}
\label{fig:fig6}
\end{figure}

Figure \ref{fig:fig6} shows double-peak and Q-like Stokes
V signals with lateral peaks formed with low intensity (darkest blue
curves) and high intensity (darkest red curves). While the former ones are
limited in amplitude, the latter ones increase with the incident intensity.

Besides, for a constant illumination along
frequency,
small opacities ($\approx 3$) create
less contrasted peaks while large opacities ($\approx 10$) create a 
well-defined central dip as a saturation effect
(Fig. \ref{fig:fig6}). Such line-core saturation renders the outgoing polarization
independent of the illumination and thus the amplitude and sign of the dip is
strongly determined by
the emissivity, through the parameter $\alpha_1$, as seen in the lower panels of
Fig. \ref{fig:fig6}. 

In other words, the
self-absorption effect producing the central dip in the polarization
profile is not increasing
with opacity nor $I_0$ once saturation is reached. As a result, the
sign and depth of the central dip 
are only determined by the balance between
emissivity and a (saturated) absorption of circular polarization.
When opacity
increases further, saturation extends to outer wavelengths, where the
peaks become 
more separated and smaller.

A nonmagnetic model
such as this first one seems unsuitable for
most anomalous solar V signals because it would require a scattering layer with unrealistically high temperatures (to fit the width of
  the profiles). Yet, the previous explanations have
  served to characterize the action of dichroism and are useful in the rest of the paper. 

\subsection{Condition of neutral medium and dichroic inversion}\label{sec:inversion}
Equation (\ref{eq:sol_rte5}) and our previous figures show that
in an optically thick atmosphere Stokes V is minimized at
frequencies in which the term depending on the emissivity of circular polarization
cancels exactly the term controlled by the illumination ($F_1=F_2$). 
In such a case we say that the
medium has become \textit{optically neutral for Stokes V} because the
action of emissivity and dichroism are in a sort of equilibrium. When this
happens only at line center in the context previously described
(with atomic orientation but without
magnetic field) the result is fully contrasted double-peak
V signals. 

Strictly speaking, Eq.(\ref{eq:sol_rte5}) shows that a neutral medium at a given wavelength
yields an outgoing polarization that is the one entering the
slab but attenuated: $V=V_0 \cdot e^{-T_{\nu}}$. Accordingly, observing a zero crossing
in Stokes V does not necessarily mean a neutral medium at that
wavelength unless the background does not produce
significant $V_0$ along the LOS
and/or the formation region is opaque enough to dampen it. 

\begin{figure}[!t]
\centering
\includegraphics[scale=0.55]{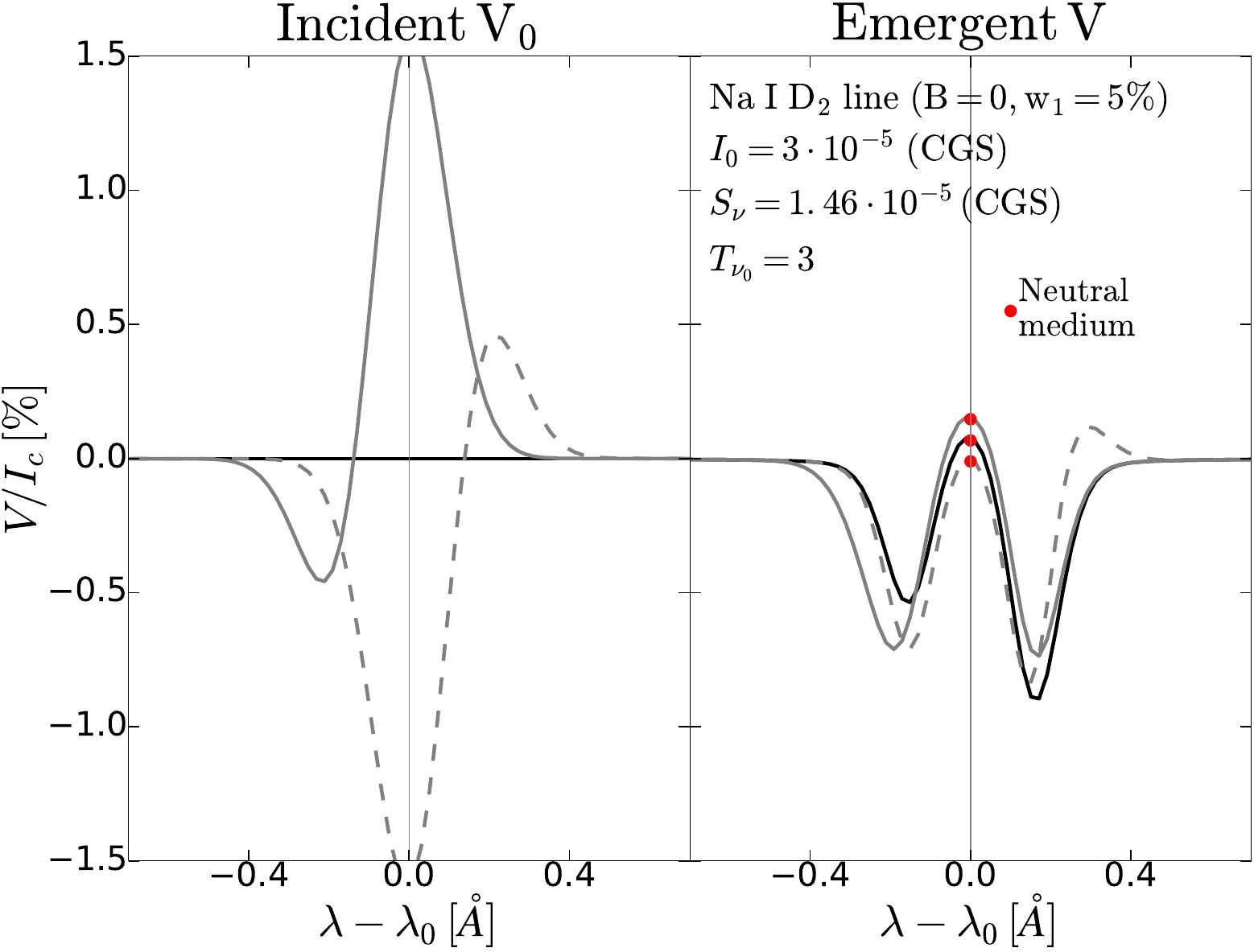}   
\caption{Effect of an asymmetric $V_0$ in a nonmagnetic slab of
  reduced opacity that is illuminated by a spectrally flat intensity.}
\label{fig:fig4}
\end{figure}

The right panel of Fig. \ref{fig:fig4} shows the emergent Stokes V with our model
for the same slab with neutral medium at line center 
but changing only the Stokes $V_0$ 
profile (left panel) at the incoming boundary of the slab. 
It is illustrative to see that, though V$(\lambda_0)\neq 0$ when
$V_0\neq 0$, the deviation
from zero is relatively small at line center (where opacity is the largest in a
slab of limited geometrical width), even for strong incoming
signals. Instead, the symmetry of the emergent signal can change
 if the medium becomes thin at other wavelengths. In the example of
the figure, we have considered that in the presence
of velocity gradients along the LOS and magnetic fields in lower layers, the slab will be
illuminated by a shifted and very asymmetric $V_0$ , as is the case for those in the
left panel of the figure. 

The condition of neutral medium for a Stokes parameter $k=$ Q, U, or V
can be written in several ways that relate the 
parameters of the model. Some insightful relations are obtained
for an atmosphere that is not optically thin ($T_{\nu}\gtrsim 3$).
For instance\footnote{The r.h.s. in this
  equation resembles a Planck function if we associate
  a ``temperature'' $T=c^2/2K\nu^2$ and an opacity $ T_{\nu} =h\nu/KT=2h\nu^3/c^2$.}:
\begin{equation}\label{eq:neutral}
 \frac{S_k-S}{I_0-S}= \frac{T_{\nu}}{e^{T_{\nu}}-1}, 
\end{equation}
with $S_k=\epsilon_k/\eta_k$. Alternatively,
\begin{equation}\label{eq:neutral3}
 \frac{S_k}{S}= \frac{1-f_0[1+T_{\nu}(1-\alpha_2)] }{1-f_0}. 
\end{equation}
We can also express this as a function of a \textit{critical intensity} $I^{\rm critic}_0$ illuminating the slab along
the LOS. For instance, for Stokes V we have 
\begin{equation}\label{eq:neutral4}
 I^{\rm critic}_0= \frac{V(I_0=0) }{\eta_V f_0},
\end{equation}
which implies that
\begin{equation}\label{eq:vv0}
 \frac{V(I_0)}{V(I_0=0)}= 1-\frac{I_0}{I^{\rm critic}_0}.
\end{equation}

Finally, the same condition written in terms of the ratio
$\alpha^{\rm critic}_2=I^{\rm critic}_0/S$ is 
\begin{equation}\label{eq:neutral2}
\alpha^{\rm critic}_2= 1+\left[\frac{S_k}{S}-1 \right]\cdot\frac{f_1}{f_0}. 
\end{equation}
This latter expression provides a useful estimation of the critical value that the $I_0$ has to cross for the sign of
the polarization to change at a given frequency (\textit{dichroic sign
  inversion}); this inversion depends on
the imbalance between emission and absorption for the Stokes parameter
considered. 

\begin{figure}[!t]
\centering
\includegraphics[scale=0.72]{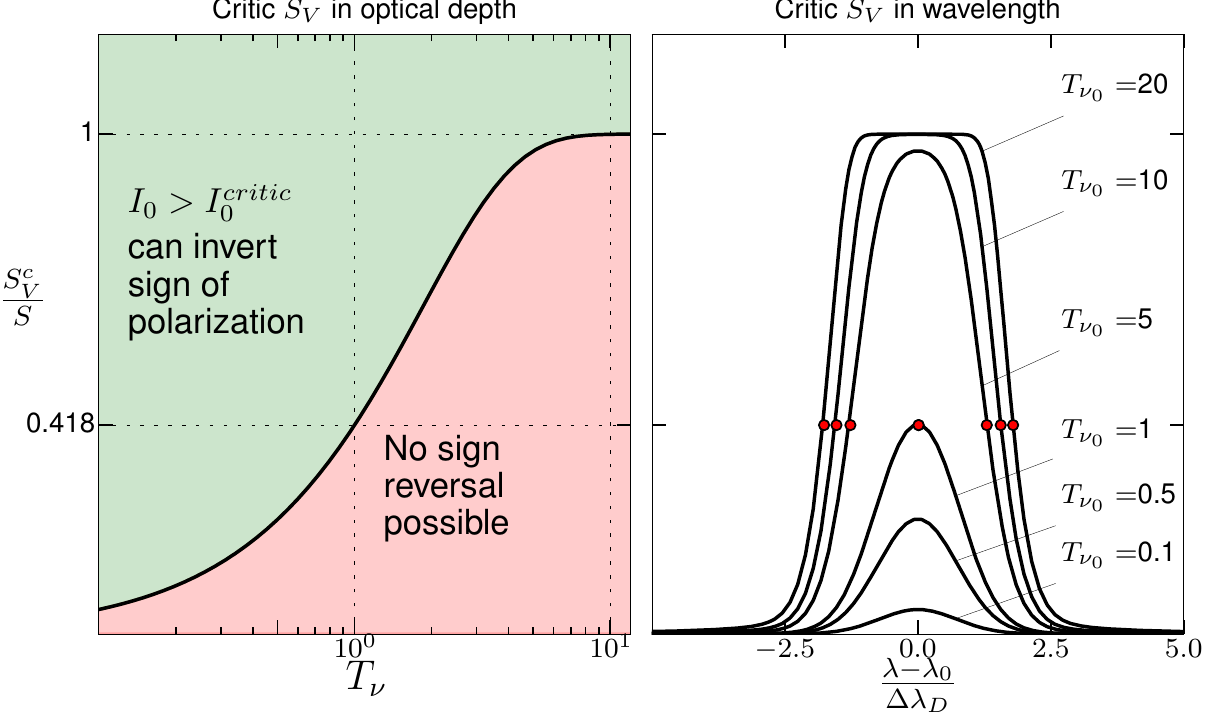}   
\caption{Left: Critical values (black line) of the quantity
  $S_V/S=(\epsilon_V/\eta_V)/(\epsilon_I/\eta_I)$. If $S_V/S>(S_V/S)^{\rm
    critic}$ (green region), then an
  $I_0>I^{\rm critic}_0$
  can modify the sign of the polarization. Right: Spectral
  representation of the same quantity for different maximum values of
  opacity in a slab. The red dots mark the positions of $T_{\nu}=1$.}
\label{fig:fig7}
\end{figure}

Here we comment on two
cases in Eq.(\ref{eq:neutral2}). The first one is
when $S_k/S=1$, which gives $\alpha^{\rm
  critic}_2=1$: in order to invert the sign of the polarization, 
$I_0$ has to jump above or to diminish below the
value of the intensity source function. A second case is when
$S_k/S=0$ (no emissivity for Stokes k), which implies $\alpha^{\rm
  critic}_2= 1-\frac{f_1}{f_0}$. This value is always negative, 
so cannot be crossed by a physical $I_0/S>0$. This
simply means that in the absence of emissivity of polarization the dichroic sign inversion is
not possible under solar conditions.
Hence, there is a minimal emissivity and a minimal critical value $S^{\rm critic}_k$, that
guarantees $\alpha^{\rm  critic}_2 \geq 0$ in Eq.(\ref{eq:neutral2}).
For Stokes V this occurs when $ \frac{S_V}{S} \geq \frac{S^{\rm critic}_V}{S} =1-\frac{f_0}{f_1}$ (see Fig. \ref{fig:fig7}).
Note that $S_V<0$ is possible but still subcritical: only
$S_V>0$ allows for a dichroic sign inversion. 

When the LOS optical depth increases at a given wavelength, 
 then $S^{\rm critic}_V \rightarrow S$. An $S_V>S$ is possible, but our preliminary tests indicate that the larger the $S^{\rm critic}_V$, the more difficult it is to have an $S_V$ above it. In other words, it
is more difficult to have $\epsilon_V/\eta_V>\epsilon_I/\eta_I$.
Thus, the probabilities of a sign inversion are lower for larger
optical depths along the LOS. 
Paying attention only to $T_{\nu}\approx 1$, where the sensitivity to the dichroic modulation is
maximum for a given frequency $\nu$ (remember Fig. \ref{fig:fig3}), the
critical value is a more feasible $S^{\rm critic}_V/S=0.418$ (Fig. \ref{fig:fig7}).

Finally, note 
that $\eta_V$ has zeros making $S_V = \infty$. This should not lead to
confusion; $S_V$ is only a derived quantity that we are using here for
convenience and compactness. When
$S_V=\epsilon_V/\eta_V=\bar{\epsilon_V}/\bar{\eta}_V$ is explicitly inserted
into Eq. (\ref{eq:new2}), $\bar{\eta}_V$ and the 
poles disappear from the denominator, always yielding finite Stokes parameters. A more detailed study
relating $S^{\rm critic}_V$ with the spectrum of the optical
coefficients and the atomic properties  
is presented in a separate publication.
\begin{figure}[!t]
\centering
\includegraphics[scale=0.7]{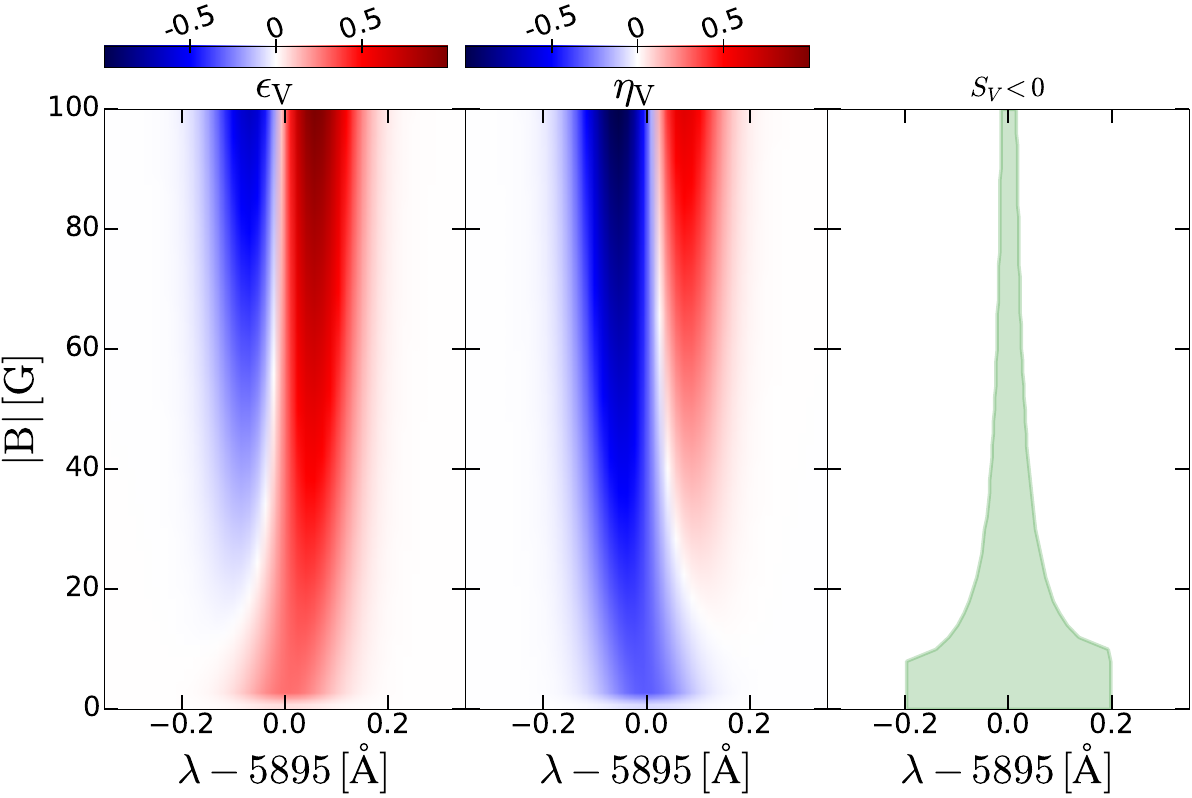}   
\caption{In green, region where $S_V<0$ due the ability of atomic
  orientation ($w^1_0=1 \%$) for dephasing the zeroes of $\epsilon_V$
  and $\eta_V$.}
\label{fig:fig8}
\end{figure}

\subsection{Condition of reinforcing medium}
We define a reinforcing medium
as one in which the polarization contribution of dichroism 
is added constructively to the one of emissivity at a given frequency. 
From Eq. (\ref{eq:sol_rte5}), this interesting situation implies
\begin{equation}\label{eq:reinforcing_media}
 \frac{\frac{S_V}{S} f_1}{(\frac{I_0}{S}-1)f_0 + f_1} < 0 \quad
 \rightarrow \quad \frac{\epsilon_V}{\eta_V}<0
.\end{equation}
The arrow follows from the fact that in the first inequality only
$S_V$ can be negative 
because S, the denominator, and the transfer
functions are always positive\footnote{Stimulated emission does
  not dominate radiative transfer in stellar atmospheres, hence
  $S\geq0$.}, regardless of the optical depth and
$I_0$. 

A superficial inspection of the equations of the optical coefficients shows
that in general the relative sign of $\epsilon_V$ and $\eta_V$ depends on the
signs of the polarizability coefficients and on the sign and amount of
atomic orientation in each level of the atomic
transition. While
the polarizability coefficients can be tabulated for different atomic
transitions \citep[see e.g., the coefficients for HFS in Table 10.7
of][]{LL04}, the sign of the orientation in each
atomic level depends in general on the geometry of the pumping radiation field
(e.g., through $J^1_0$) and
on the detailed solution of the SEE. 

Considering for a
moment a weak magnetic
field along the LOS, and in the particular case of the Na {\sc i} D$_1$
line optically pumped without radiation field orientation, we find
that both $\epsilon_V$ and $\eta_V$  always have the same sign,
frequency by frequency, meaning that no reinforcing medium is possible. This is true even 
though their signs change together
with frequency due to the
Zeeman splitting. We then find that the only requirement for having
a frequency interval with opposite signs in $\epsilon_V$ and $\eta_V$
(hence, a reinforcing medium) is a ``dephase'' between the
(Zeeman-induced) zero crossings of both optical coefficients. 
An analysis of optical coefficients including the physics of
scattering polarization 
shows that in Stokes V the dephase can be simply achieved with atomic
orientation, as shown in Fig. \ref{fig:fig8}. Thus, atomic
orientation is able to open spectral windows where the polarization
 amplitude could be reinforced. The width of the window decreases with the
 magnetic field strength but increases with the atomic
 orientation. As such windows are around the zeroes
of the optical coefficients, the reinforcement of the amplitude by
this mechanism seems in general limited.

\begin{figure}[!t]
\centering
\includegraphics[scale=1.2]{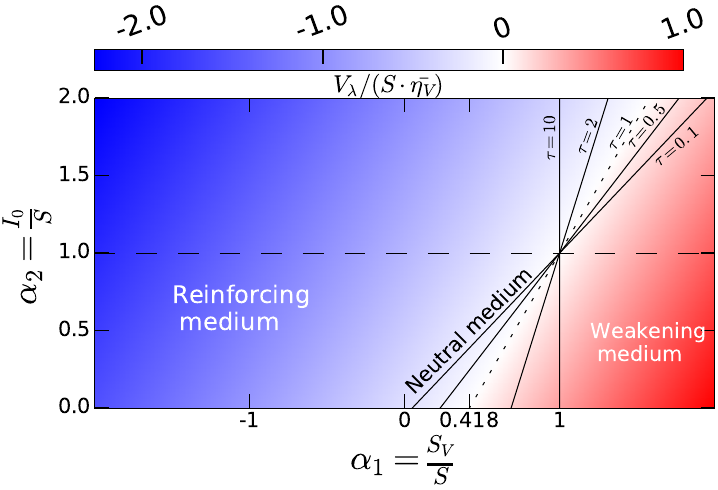}   
\caption{The quantity $V_{\lambda}/[S\cdot \bar{\eta}_V$] shows (in color) the
  dependence for any wavelength and spectra line of the emergent Stokes V amplitude on the backlight
  illumination (vertical axis) and on the source function for V (horizontal axis) for
  an atmosphere with $T_{\nu_0}=1$. The
  medium becomes neutral along the dashed line. For other optical
  depths, the background colors remain qualitatively similar but the
  neutral medium takes place at the corresponding continuous lines.}
\label{fig:fig9}
\end{figure}
Figure \ref{fig:fig9} summarizes the concepts of neutral medium,
reinforcing medium, and dichroic inversion. When $\alpha_1<0$ the
medium is reinforcing and the
illumination cannot change the sign of the emergent 
polarization at a given wavelength, which is always the sign given by emissivity. When
$\alpha_1>0$ the illumination existing at a given optical depth 
can induce a dichroic sign inversion if $\alpha_2$ crosses the critical
value delimited by the (black) lines in the figure. We advance
that a Doppler brightening produced by velocity gradients
along the LOS can produce first a cancelation of
the signal and then enter in the regime of reinforcing media, in
which larger intensities produce larger signals with an opposite sign 
to the one in the initial static situation.

\section{Dichroism with Zeeman splitting}\label{sec:magcase} 
In the following, we investigate the mechanism by which the Stokes V signals are
generated in the presence of
magnetic fields, NLTE, and atomic polarization.
We have extended the model
of previous sections to consider Zeeman splitting, a spectral-line profile for
the illumination $I_0$,
and a relative LOS macroscopic motion between lower layers (those
generating $I_0$) and upper layers (main scattering region). Equation (\ref{eq:new2}), or equivalently
Eq. (\ref{eq:sol_rte5}), still applies and forms the basis of the model
but now the optical coefficients must additionally account for
magnetic effects such as the Zeeman splitting.
In order to describe the radiative transfer for a generic spectral
line, the temperature (T), the LOS velocity ($\upsilon^{\rm los}$), the
magnetic field (B) of the slab, and the minimum intensity are
parameterized through the generalized variables:
\begin{subequations}\label{eq:parameters}
\begin{align}
\alpha=& \frac{w}{\Delta\lambda_D}, \\
\beta=& \frac{1}{2}\frac{\Delta \lambda_{\rm
    B}}{\Delta\lambda_D}\bar{g}_L=\frac{1}{2}\frac{(\lambda^2_0/c)\nu_{\rm
    L}\bar{g}_L}{\Delta\lambda_D}, \\
\xi=&\frac{(\lambda_0/c)v^{\rm los}}{\Delta \lambda_D},\\
\delta=& I_0/S - 1, \quad ( \rm at \, \lambda-\lambda_0=0),  
\end{align}
\end{subequations}
where $\Delta\lambda_D$ is the thermal width of the absorption
profiles at the scattering layer, $\Delta \lambda_{\rm
    B}$ is the corresponding Zeeman
splitting, $\bar{g}_L$ is the effective Landé factor associated to the
transition, $\nu_{\rm
  L} [\rm s^{-1}]=1.3996\cdot 10^6 \cdot \rm{B_{LOS}[G]}$ is the Larmor
frequency of the longitudinal magnetic field component, and $w$ is the spectral
width at half depression of the incident line profile $I_0$ (or at
half amplitude if $I_0$ is in emission). 
The factor $\alpha$ quantifies the relative width among the
background intensity profile and the absorption coefficient.
 We have also defined the \textit{relative darkening} $\delta$ (defined as $<0$
when $I_0<S$), to be used in the following section.

The parameters $\alpha$, $\beta,$ and $\xi$ are in units of $\Delta\lambda_D$. 
Expressing $\xi$ in units of $w$ is just a matter of dividing 
by $\alpha$. In general, one may also work in units of the total FWHM
of the $\eta_I$ of the
scattering layer by multiplying $\alpha$ and $\xi$ 
by\footnote{We relate the Doppler width of a 
Gaussian profile to its FWHM with FWHM$=2\sqrt{\ln{2}}\cdot
  \Delta\lambda_D$ and we add the magnetic broadening with $\beta$.} $3/(5+6\beta)$, or simply by $3/5$ if the magnetic
broadening is negligible. The idea is to include as many broadening
mechanisms\footnote{There are seven broadening mechanisms: magnetic,
  thermal, microturbulent, collisional, opacity saturation, and kinetic
  (the latter two due to lack of resolution in depth along the
  LOS). The seventh one is instrumental, due to
  lack of spatiotemporal resolution.} as possible in these relative definitions between the
layers, but in this paper we simplify by only using variables
in Doppler units.
\begin{figure}[!t]
\centering
\includegraphics[scale=0.35]{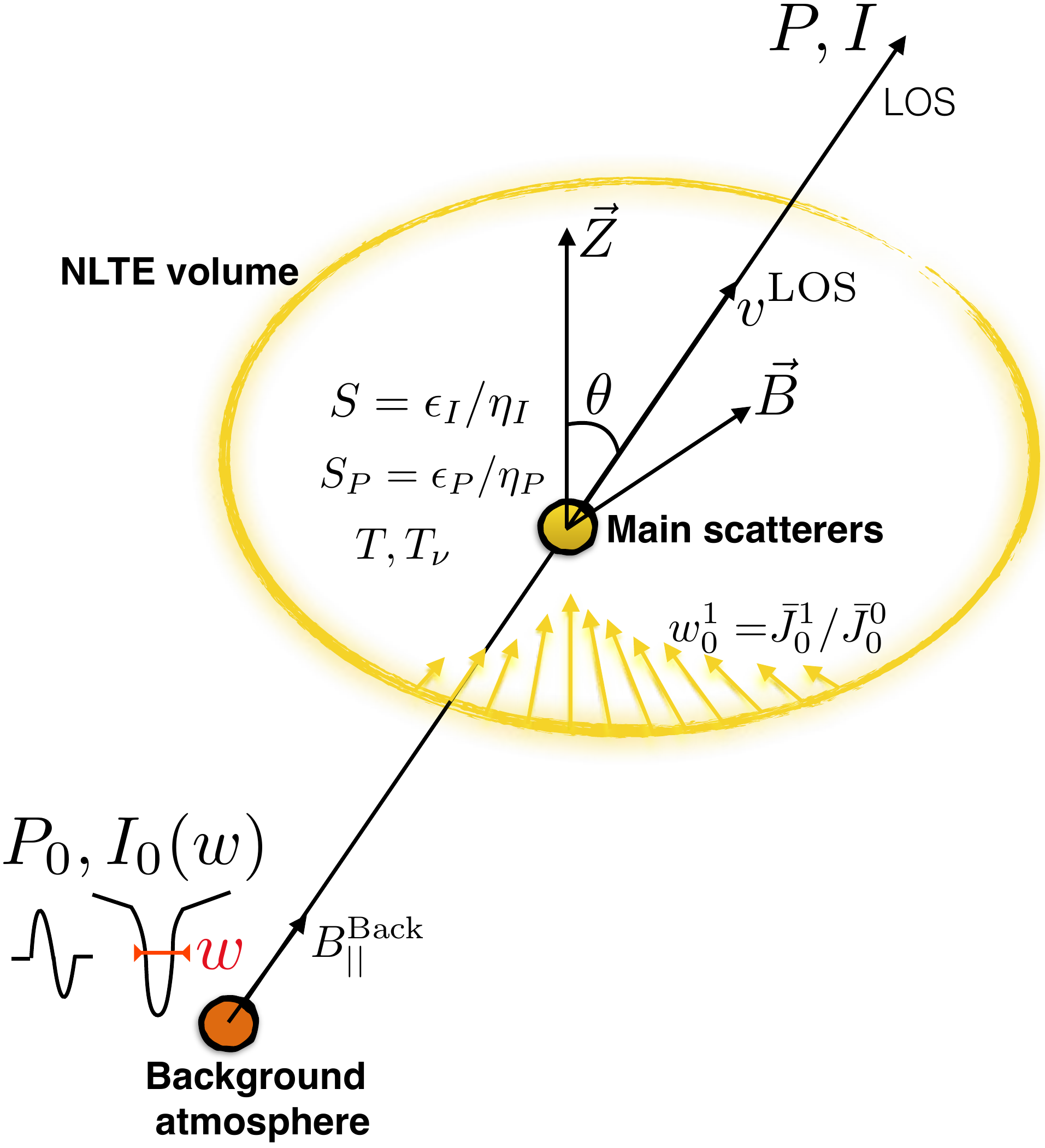}   
\caption{Scheme of the first version of our model for polarization.}
\label{fig:fig11}
\end{figure}

Regarding the velocity, what is important to explain the 
signals is the velocity gradient between the main and the
  lower (background) layers. Hence, the analysis is simplified by setting the velocity of the
lower layers to that of the laboratory frame (i.e., to zero) and
measuring the slab velocity with respect to this latter (velocity is positive when towards the observer). Thus, $\xi$ is a measure of both velocity and velocity
gradient.
\begin{figure*}[t!]
\centering$
\begin{array}{cccc}
\includegraphics[scale=0.73]{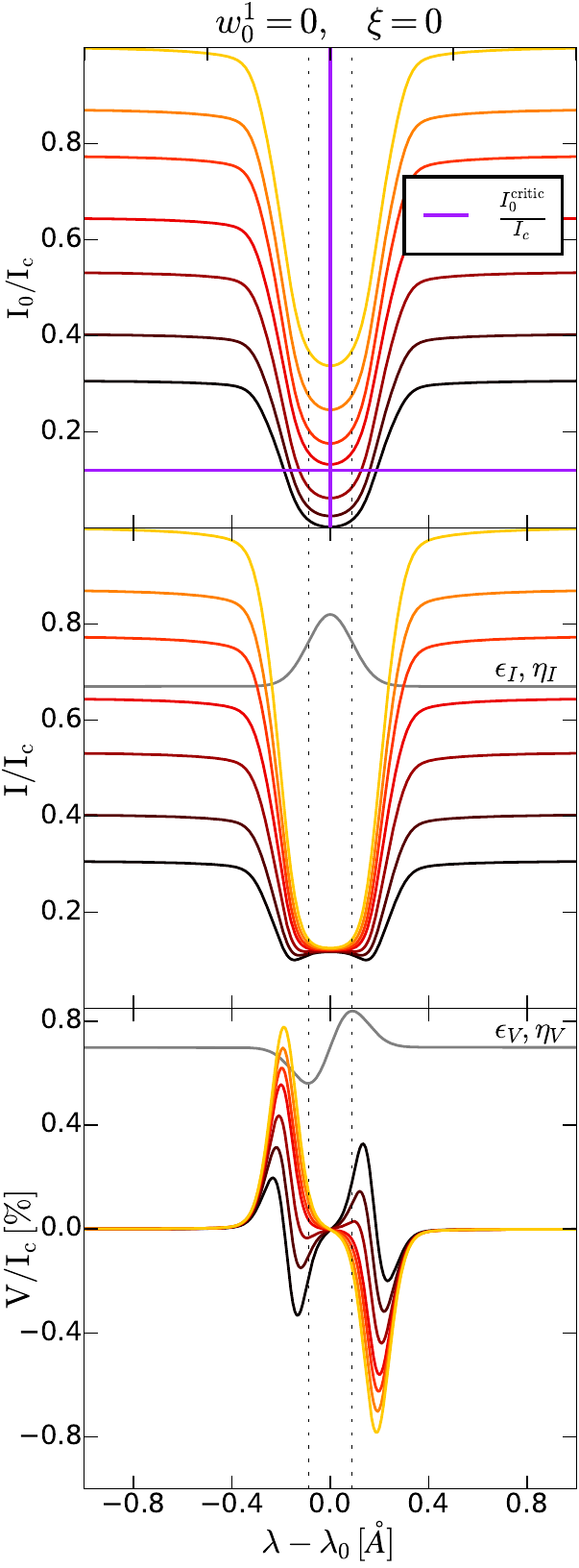}    
\includegraphics[scale=0.73]{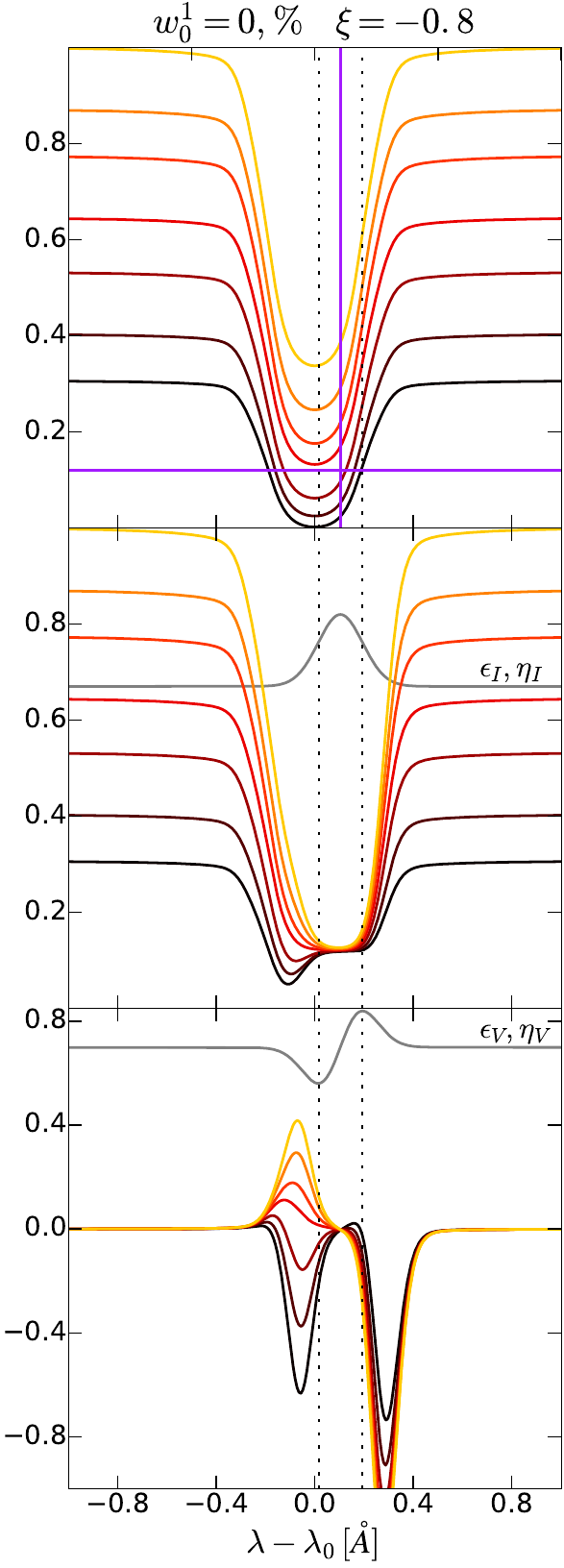}   
\includegraphics[scale=0.73]{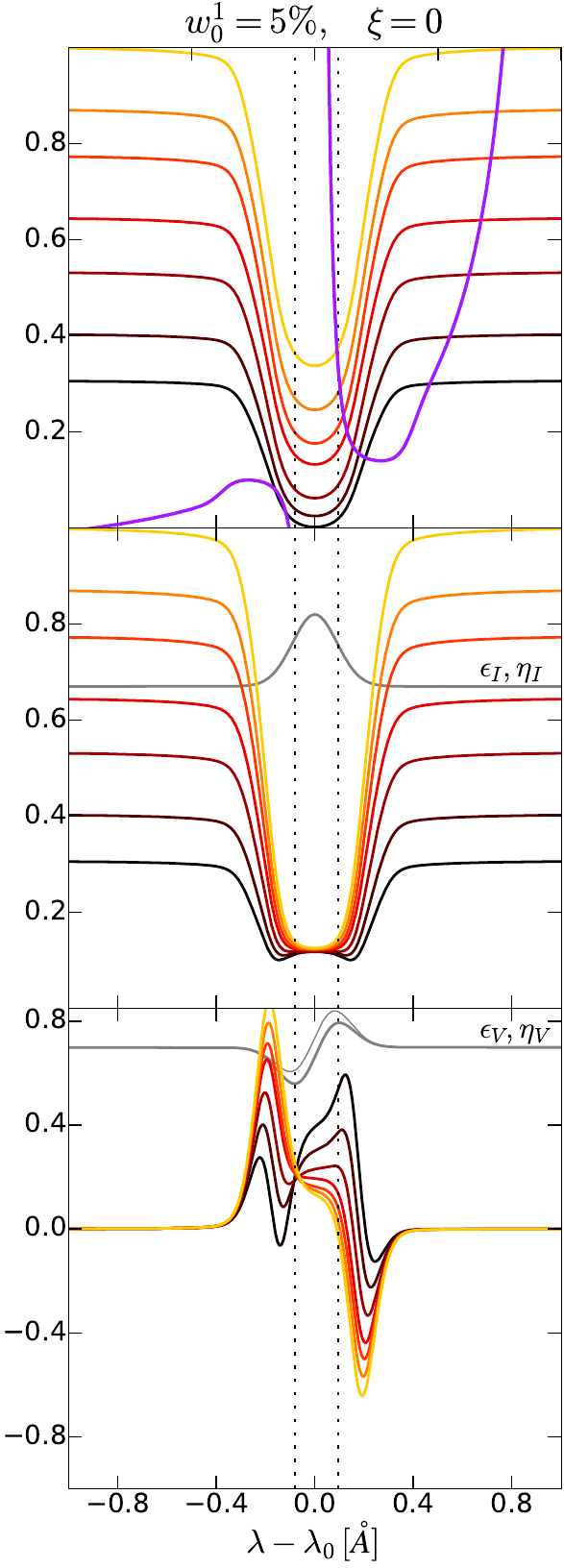}   
\includegraphics[scale=0.73]{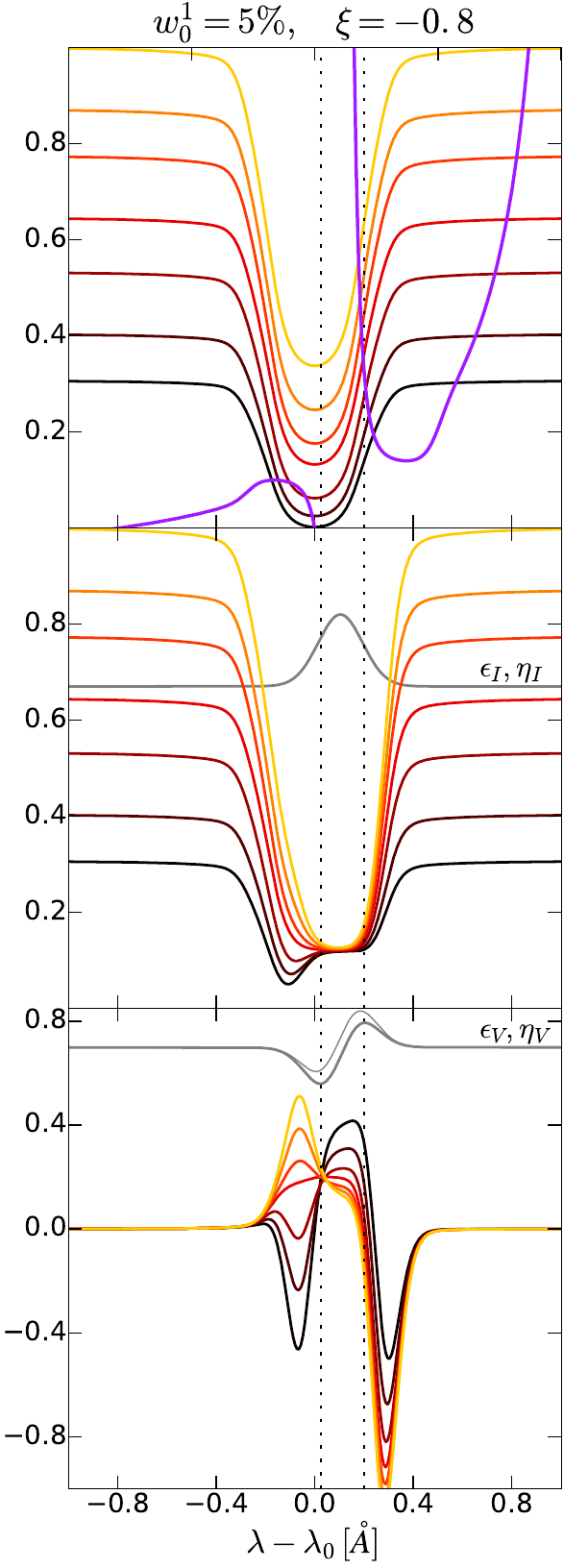}   
\end{array}$
\caption{Formation of NLTE circular
  polarization with atomic polarization and velocity gradients in the Na {\sc i} D$_1$ line. The figure indirectly explains why the double-peak and Q-like V signals
of the solar Na {\sc i} D lines are more often found near the solar limb (black
profiles) than at the disk center (yellow). The calculation is done with $\mu=0.1$, $T_{\nu_0}=4 $, $\alpha=3.14$
  (with $T=6.5$ kK and microturbulent velocity $\upsilon_{\rm micro}=6.5 \, \rm km
  \,s^{-1}$),  $I_c=2\cdot
  10^{-5}$ (cgs units), $\beta=0.01$ (B=$154$ G with $ \vec{B}$ parallel to
  LOS vector). Violet lines: critical intensity $\alpha^{\mathrm{critic}}_2 =I^{\mathrm{critic}}_0/S$
}\label{fig:fig10}
\end{figure*}
The effect of velocity gradients is two-fold: On one hand, they
introduce a Doppler spectral shift between the pumping field and the
absorption profiles of the main formation region. Velocity gradients
along the LOS thus produce Doppler brightenings at such heights,
increasing $I_0$ and bringing the possibility of a dichroic sign
reversal. On the other hand, they increase the amount of
atomic orientation by affecting the contribution of the radiation field
to the rate equations. 
To consider this latter effect in the following numerical
experiments, we have set an ad-hoc fractional radiation field
orientation $w_1=5 \%$ 
for both $D_1$ and $D_2$ transitions of Na {\sc  i}.
 Following the approach in HAZEL \citep{Asensio-Ramos:2008aa}, we calculated the rest of the
radiation field tensors by using the angular and frequency
dependence of the photospheric solar continuum radiation tabulated
by \cite{Pierce:2000}. With the information on the radiation tensor, the atomic density
matrix is calculated by solving the SEE in NLTE.  
 However, for calculating the emergent Stokes vector we did not use
the average values of \cite{Pierce:2000} in the LOS lower-boundary
condition of the RTE. Instead, we introduced ad-hoc
intensity profiles and thus investigate the dependence of the emergent
circular polarization to the LOS illumination. This change significantly increases
the diagnostic capability of HAZEL without the need of multi-layer radiative transfer.

\subsection{Detailed explanation of the formation of Stokes V anomalies}\label{detail}
We can now explain some basic results of our model. Figure \ref{fig:fig10} shows ad-hoc
background intensity profiles (upper panels) with the corresponding emergent
intensities (middle panels) and emergent circular polarization
(lower panels) calculated in four cases of interest (one per column) for the Na
{\sc i} D$_1$ line.
 The only parameters of the experiments that change among columns
are the LOS Doppler velocity $\xi$
and the radiation field orientation $w^1_0$. To ease the
  explanation, in this section the background polarization
$V_0$ is set to zero.

The intensity profiles in the upper
panels are symmetric and
centered on $\lambda_0$, with different minimum 
(i.e., line center) amplitudes $I_0(\lambda_0)/S$ but the same spectral
width $w$, chosen to resemble the width of the observed solar profiles of
this spectral line \citep{Stenflo:2000, Stenflo:2001aa}. 
The violet line here is the critical intensity producing a neutral medium, that is, the threshold
marking the dichroic sign inversion of the
emergent circular polarization. The only difference between
each upper panel is that the critical intensity is no longer
spectrally flat when atomic orientation is present in the scattering layer
(i.e., when $S_V$ departs from
$S$; see Eq. (\ref{eq:neutral2}) and the following sections for details).

The optical coefficients for absorption and emission (gray lines in middle
and lower panels) are plotted with arbitrary
amplitudes, simply to
have a reference of their spectral shape; they have a total broadening
that is smaller than the total width of the background $I_0$. 
Note that when the total opacity $T_{\nu_0}$ of the scattering region along
the LOS is above one (optically thick plasma), the 
width of the intensity spectrum does not correspond to the thermal width.
In these numerical experiments, $T_{\nu_0}=4$,
and then opacity saturates the profile enhancing its spectral width
 beyond the thermal value. This well-known effect is
spectrally identical to that shown for $S_V$ in right panel of Fig. \ref{fig:fig7}. 

In all cases of Fig. \ref{fig:fig10} the emerging
 intensity  
follows other well-known behavior: its value tends to the source function of the
 slab, thus producing an absorption or an emission 
for incident intensities above or below $S$,
respectively. In this sense, the four emergent intensities only
differ in that when 
a Doppler shift is considered (second and fourth columns), the optical
profiles are shifted with respect to the illumination and consequently the part of the
intensity profile that is affected by it is also shifted. Here this effect
is referred to as kinetic broadening.

Let us analyze Figure \ref{fig:fig10}  column by column. The first column corresponds to a situation
 with no atomic polarization or Doppler shifts.  
Observing the upper and lower panels of this column, we see that 
in wavelengths where the absorption profile captures subcritical incident
intensities (darkest profiles) 
 the emergent Stokes V develops additional central peaks whose 
 signs are opposite to those in any other wavelength and/or
 cases in which the incident intensity is above the critical
 value. 
 Without velocity gradients or atomic orientation, 
the sign reversal of such central peaks is symmetric around line center, and consequently the emergent CPs remain anti-symmetric. As explained in previous sections, the sign reversal is a result of
the competition between dichroism and emission when the source
function for Stokes V is positive.
 When the whole $I_0$ captured by $\eta_V$ is supercritical (lines with yellowish-warm colors), the V profiles
can only have two peaks. The same would happen if in the absorption
band the intensity profile
were fully below the critical value.

The second column of panels in the figure illustrates the pure effect of a velocity
gradient between the lower atmospheric layers and the region of
formation of the emergent Stokes V. The relative Doppler shift between
$I_0$ and the absorption profile causes the absorbing atoms to capture more
light of the line wing (Doppler brightening) and increases 
the spectral asymmetry of the background intensity between the peaks of $\eta_V$. 
Such spectral asymmetry would be
maximized for all the intensity profiles when the Doppler shift is
$|\Delta \lambda_D \cdot 5/3|$ (i.e., $|\xi|= 5/3 $). In this figure we 
arrive only to $|\xi|= 0.8 $, and only for the darkest profiles
does the Doppler shift approach a crosspoint between
the critical intensity and $I_0$ (neutral medium). Thus, intensities ranging from
supercritical to subcritical values (only in the blue half of the profiles,
where also S$_V>S^{\rm  critic}$)  switch
that half of V/I from positive to negative values,
changing the profile from antisymmetric to a signal with only one
sign, thus giving a double-peak profile. 

One could say that the red half of the Stokes V spectra in the
bottom panel of the second column is {dominated} by
dichroism, 
while the blue halves of the two
cases with the weakest intensities are dominated by emission of
circular polarization. However, this could be misleading because it is actually
the balance between emission  and absorption that is important: if any of the two
contributions disappear, the anomalous signals cannot be explained with Zeeman splitting.
Thus, as the background (say photospheric) intensity is significantly
stronger near disk center than near the limb, the
disk center signals can be
more often dominated by dichroism, while exhibiting wavelengths 
dominated by emission near the limb.   
This can be counter-intuitive, but we remark that 
the variations of Stokes V at the red lobes in our example are all dominated by
dichroism because $I_0$ is always
supercritical and opacity is still significant. It is in the anomalous lobe
(the blue one) where the dichroic contribution tends to zero when $I_0$ is low.

Subsequently, the rule in
this kind of situation is that the sign
reversal (with respect to a situation with supercritical $I_0$) 
 happens in the lobe receiving less background intensity during the
 Doppler shift. 
Obviously, as an $I_0$ that is
completely subcritical produces
antisymmetric V profiles of opposite sign to those for an
intensity fully above the threshold, the
polarity of the magnetic field cannot be unequivocally deduced
without knowing the $I_0$
that illuminates the scatterers. While in intensity a strong reduction
of absorption is easily identified as an emission line, in the case of
nonanomalous Stokes V signals it is more difficult to decipher because the signal
stays antisymmetric and only changes its sign.
This sort of ambiguity between magnetic field polarity and
an unknown level of background intensity can be avoided in
double-peak profiles (and other anomalous signals) if the sign
of the relative Doppler shift between different layers is known. 

The third column of Figure
\ref{fig:fig10} shows the result of including a positive radiation
field orientation in our numerical experiments. 
The critical intensity now exhibits two separated
spectral branches
due to the presence of a zero in the absorption coefficient $\eta_V $ that is
not balanced by a zero in $\epsilon_V$. 
Thus, at wavelengths where $\eta_V=0$
the critical intensity presents an asymptote and 
the amplitude of the Stokes V signals becomes
 independent of $I_0$, which explains why all the Stokes V profiles
 coincide at that point. There, Stokes V is only due to emissivity, 
and has a nonzero amplitude (a
central bump) if there is
 atomic orientation in the upper level. This could be used as an
 observational diagnosis: if we could
 identify where $\eta_V=0$, we could calibrate
 and measure the level of
the upper-level atomic orientation at that wavelength.
We note that the Stokes V
 profiles represented with a yellowish line in the third column of
 Fig. \ref{fig:fig10}, with such a
 particular enhancement of the line core signal and area asymmetry, 
have always been very common in observations of the solar photosphere \citep[see
 e.g.,][]{Rueedi:1992aa}. 

\begin{figure}[h!]
\centering
\includegraphics[scale=1.2]{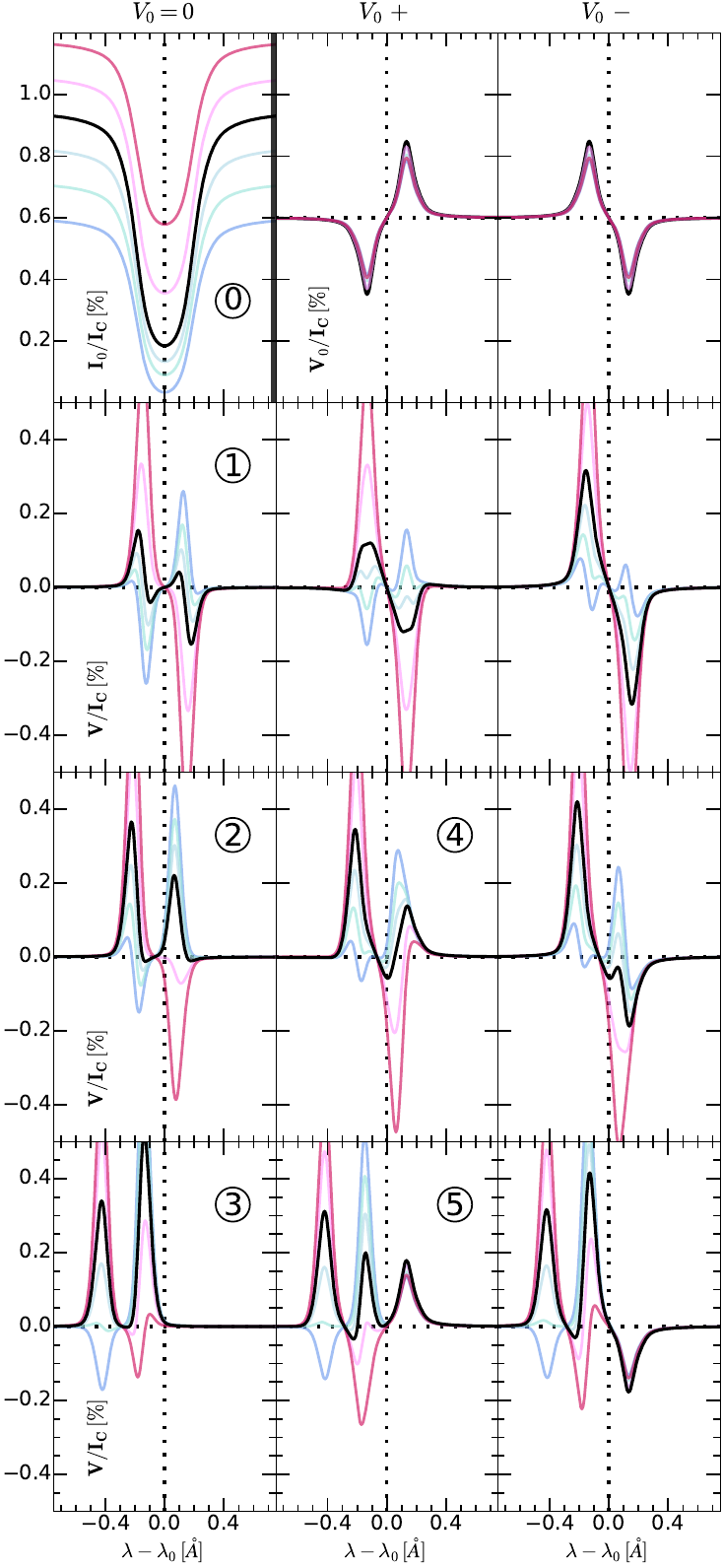}  
\caption{Variation of Stokes V in Na {\sc
    i} D$_1$ when changing $V_0$ (see column titles)
 assuming $\xi=1.3, 0.3$, and $0$ for bottom, second, and third rows,
respectively. Other parameters are $T_{\nu_0}=5$, $B=100$
G, $w^1_0=0$, $I_c=2\cdot
  10^{-5}$ (cgs units), and $\alpha=1$. The background radiation is represented in
  the top row, with $V_0$ calculated from $I_0$ using the weak-field
  approximation. The $V/I_c$ and $V_0/I_c$ profiles have the same vertical scale.}
\label{fig:fig18}
\end{figure}
Including a velocity gradient to the calculation with atomic
orientation, we obtain the result of the fourth column in
Fig. \ref{fig:fig10}. Now the intensity differences between the red and
blue sides are again maximized, as in the second column, and
a central bump in V/I appears, as in the third column,
resulting in a double-peak profile with a central bump of inverted
sign and an amplitude comparable with the other peaks, that is, a Q-like profile. 
These results show that area and
  amplitude asymmetries in Stokes V can be created by atomic
  orientation, which thus contributes to explain Q-like Stokes V profiles.

 However, a point to note is that with or without atomic orientation, the
 Q-like profiles that one can
 synthesize within the framework of Fig. \ref{fig:fig10} will not
 exactly match the observations because the distance between 
external peaks of the synthetic Q-like Stokes V profiles is
unrealistically short. Effectively, we have observed (Carlin et. al, in prep.) that in
chromospheric lines such as the Na {\sc I} D lines the distance
is always larger than that predicted by the
weak-field approximation, meaning that the two peaks of the same sign 
shown in our figure should be significantly more separated, lying at
both sides of the intensity line core to match the observations. 
Figure \ref{fig:fig10} shows that the velocity gradient
increases that distance, but it proportionally increases the width of the corresponding
intensity profile as well, and therefore the problem persists. Thus, in order to complete our
explanation additional elements are needed.
 
\subsection{Curling the curl:  the NLTE source function and the
  background circular
  polarization}  
To complete the explanation of Q-like profiles, and thus
  assure that we can explain other polarization signals with our
  model, we have to include two additional facts. 
Let us start by considering a chromospheric line observed near the disk center and
forming in a magnetic field whose polarity does not change along the LOS.
The first fact to examine is the behavior of the 
 NLTE source function which, once decoupled from Planck, can be strongly and 
systematically decreased
 with respect to its value in the layers immediately below.
On one hand, this decreases the line-core critical intensity triggering
 a dichroic sign inversion at the scattering layer, as indicated by Eq. \ref{eq:neutral2}.
On the other hand, it furthermore increases the ratio $I_0/S$
via the denominator (and not only via $I_0$ as in Fig. \ref{fig:fig10}). 
As the source function is modified along the LOS
(for instance when it tries to re-couple with the Planck function
 during the emergence of shock waves), the relation between
 $I_0$, $S,$ and  $I^{\rm critic}_0$ changes. With the help of more
 realistic simulations we are finding that 
such an interplay is predictable and defines interesting situations. One
 of them occurs when the intensity line-core goes
 from supercritical to subcritical (or viceversa) across the {whole}
absorption band, as illustrated by panels
 0 and 1 in Fig.~\ref{fig:fig18} using our model. Here we see that radiation is 
inverting the Stokes V polarity
imposed by the magnetic field, hence we talk about a \textit{radiative
  polarity}\footnote{Stokes V polarity = magnetic polarity x
    radiative polarity}. This total reversal can happen with and without velocity
  gradients, depending on the values of $\xi$, $\alpha$, S, and $T_{\nu}$.

The second point to be
explained connects the first point with the background $V_0$. 
While lower layers contribute with a
Stokes V component of a given polarity ($V_0$), the upper layers,
even when having the same magnetic polarity, 
can contribute with an opposite radiative polarity (inverted with respect to $V_0$).
 
When combining
these two situations with the \textit{partial} dichroic inversion induced only
in one peak by mild velocity gradients (as explained in Fig. \ref{fig:fig10}), 
 different combinations of
signs and amplitudes are possible in the transfer of $V_0$ through the
scattering layer, as Fig.~\ref{fig:fig18} demonstrates. In this
figure, all $V/I_c$ panels of a given row correspond to
identical physical situations in the scattering layer, in particular
having the same magnetic field orientation and with a
source function that allows \textit{total static} dichroic inversions
 when $I_0$ changes, as illustrated. However, they
differ in $V_0$ by columns: when the longitudinal magnetic field vectors in
the foreground and background layers are either parallel or antiparallel, one
obtains the profiles in the right-most or middle columns,
respectively. When there is only one magnetic layer
 ($V_0=0$, first column), the lateral peaks of Stokes V have both incorrect locations and
spectral separations in relation to the
intensity profile and to what is observed in the sun. On the contrary, when the
background magnetic field is included, the
separation between the peaks increases with the velocity gradient, 
enclosing the intensity core beyond the maximum of its derivative with
wavelength (weak-field approx.), as required. Compare for instance
black profiles 2 and 4 for small velocity gradient, or 3 and 5 for
relatively large velocity gradient in Fig. \ref{fig:fig18}. Thus, combinations of magnetic
polarities with velocity gradients develop anomalous CP with distinctive
 morphologies. Note the triple-peak profile (three peaks of the same sign) in panel 5: 
it cannot be reproduced with a single magnetic layer.

These results imply that eventual reversals of magnetic polarity
that may exist along the LOS (as represented in the middle
column of Fig.~\ref{fig:fig18}) can play a role in the formation of some of the 
anomalous CP signals. A more remarkable
implication is that multiple unresolved components are not needed to
explain the anomalies. This is however a very 
common approach in the literature. The point is that there is another dimension where
resolution matters, and this is the depth along the LOS. Paradoxically,
we have deduced the consequences of this fact by reducing the resolution of the
radiative transfer along the LOS to a minimal number of two
layers.

As an unavoidable consequence of NLTE, the LOS resolution
is linked with the variations of magnetic polarity by means of the
geometrical extension of the formation
region: the more expanded, the more variations (magnetic and
dynamical) can it contain, and the more
LOS resolution is needed.
As pointed out in \cite{Carlin:2016aa}, an expanded
formation region can cover heights with significantly different magnetic
field orientations, which in those simulations produced anomalous Q and U signals with different
asymmetric spectral profiles in the same pixel.

\subsection{Example of application: modeling anomalous
  polarization in Fe {\sc i} in $1.5$ $\mu m$ at solar intergranules}
\begin{figure}[h!]
\centering
$
\begin{array}{ccc}
\hspace*{\fill}%
\includegraphics[scale=0.79]{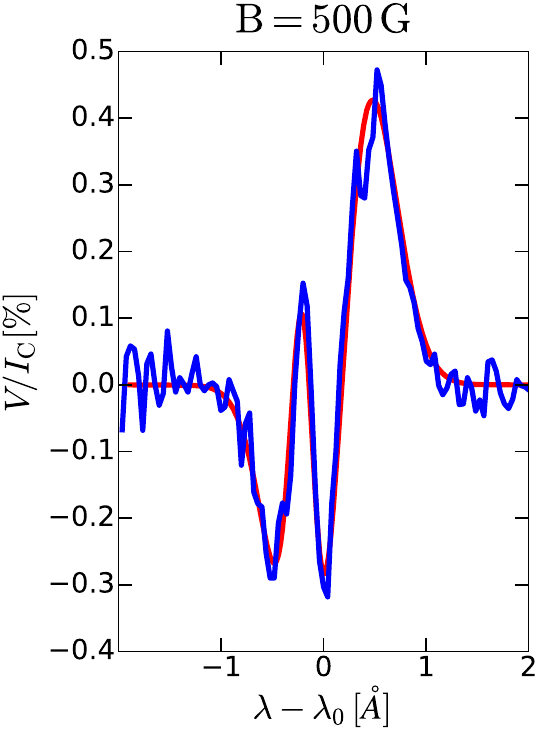}\hfill
\includegraphics[scale=0.79]{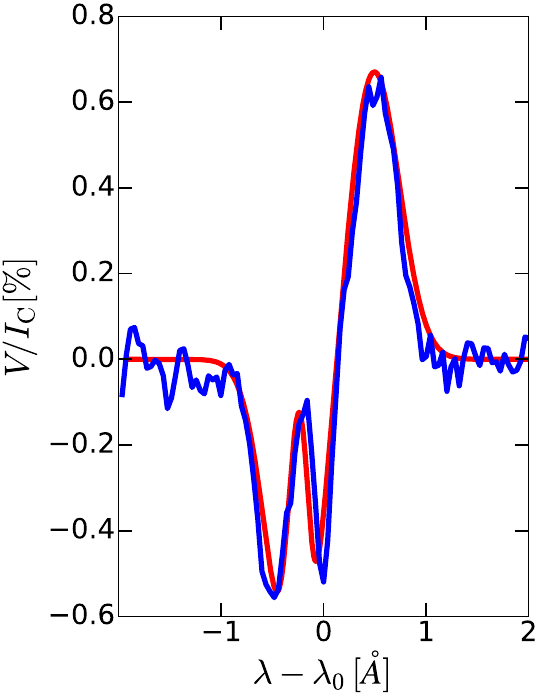}  \\  
\hspace*{\fill}%
\end{array}$
\caption{ Blue profiles: Observations of the Fe {\sc i} $15648.52$
  {\AA} (air wavelength)
  line in the quiet sun \citep[courtesy of Dr. J.M. Borrero; see
][]{Kiess:2018aa}. 
Red profiles: NLTE synthesis obtained with our model 
assuming a two-level atom for the transition $e^7D_1$ - $n^7D^o_1$,
with $A_{u{\ell}}=1.94 \cdot 10^{-6}$, effective Landé factor $g^{\rm
  eff}_{L}=3$, negligible damping constant $a=0$, $T=8000$ K, and
$w^1_0=-0.03$. The magnetic field and the LOS are vertical. 
For obtaining profile $1$
(left panel): $B_{LOS}=500$ G, $T_{\nu_0}=1.0$, $v^{LOS}=-6.3 \,
km\,s^{-1}$, $I^{\rm cont}_0/S^{\rm
  line}=1.37$, $I^{\rm core}_0/S^{\rm
  line}=0.52$, $r_c=12$ (ratio continuum to line opacity). 
For the profile $2$ (right panel): $B_{LOS}=1100$ G, $T_{\nu_0}=0.35$,
$v^{LOS}=-3.2 \, km\,s^{-1}$, $S_{\rm cont}/S_{\rm line}=1.37$,
 $I^{core}_0/S=0.93$, $r_c=18$.
The $I_0$ was obtained by using the shape of the observed $I$ as seed
profile and scaling it with $I^{\rm cont}_0/S^{\rm
  line}$ and $I^{\rm core}_0/S^{\rm
  line}$ to reproduce simultaneously the observed
intensity and Stokes V profiles within a tolerance of $5\%$ of the
continuum intensity.}
\label{fig:fig16}
\end{figure}

Observing quiet-sun intergranular lanes, \cite{Kiess:2018aa} found near-ubiquituous three-lobe Stokes V
  signals in the $1564.8$ nm Fe {\sc i} line. Carrying out detailed modeling 
and inversions, these authors concluded that
two magnetic field components with very different
strengths ($9$ G and $1147$ G) coexisting in the same
pixel are required to explain the signals. We
synthesized two profiles of their dataset with our model assuming
that the signals are instead spatiotemporally resolved. Following a 
two-level atom approach our preliminary results indicate
 that the whole Stokes vector can be approximately reproduced by two different
scenarios: 1) with a single magnetic layer ($V_0=0$) the modeling in
Fig. \ref{fig:fig16} (read caption)
is obtained if $3\%$ radiation field orientation is allowed;
and 2) with two consecutive magnetic layers along the LOS and without
atomic orientation, 
three-peak profiles can be reproduced as done for Sodium D$_1$ in the
bottom-right panel in Fig. \ref{fig:fig18}.

Due to the particularities
of the Fe {\sc i} line, more work is required to
understand the curious implications of these 
physical scenarios, and which of the two corresponds to the
observation. 
In addition, we need to
reduce the differences with the observation (up to $5\%$ of the continuum
intensity) in the wings of the synthetic intensity
 profile (we note that the signals are ``inverted'' manually,
without automatic error minimization). 
Despite this, we think that the preliminar results support our model 
as a good starting point for explaining the signals without assuming
spatiotemporally unresolved components.

\begin{figure*}[h!]
\centering
\includegraphics[scale=0.8]{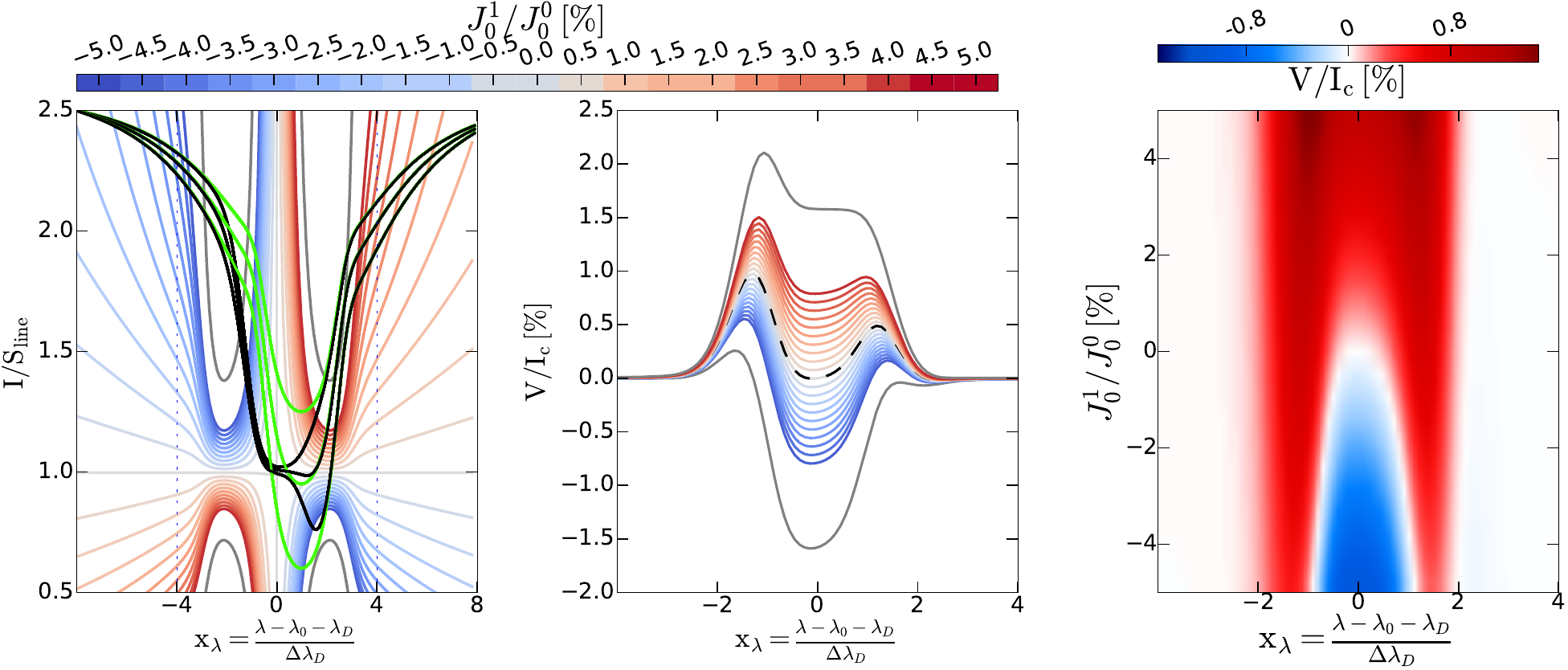}  
\caption{Examples of critical intensity in Na
  {\sc i} D$_1$. \textbf{Left panel:} Three
  background intensities (green lines) and their corresponding
  emergent intensities (black lines) on top of the critical
  intensity spectra computed for different values of atomic orientation (see
  discrete color bar) and normalized to $S_{\rm line}=min(S)$. The gray lines are references corresponding to $w^1_0=10\%$. \textbf{Middle panel:} Stokes V
  signals corresponding to the deepest intensity profile. The case
  $w^1_0=0$ is made more visible by the dashed black line. \textbf{ Right
    panel:} Same signals as in middle panel but plotted in 2D.}
\label{fig:fig12}
\end{figure*}

\begin{figure*}[t!]
\centering
\includegraphics[scale=0.6]{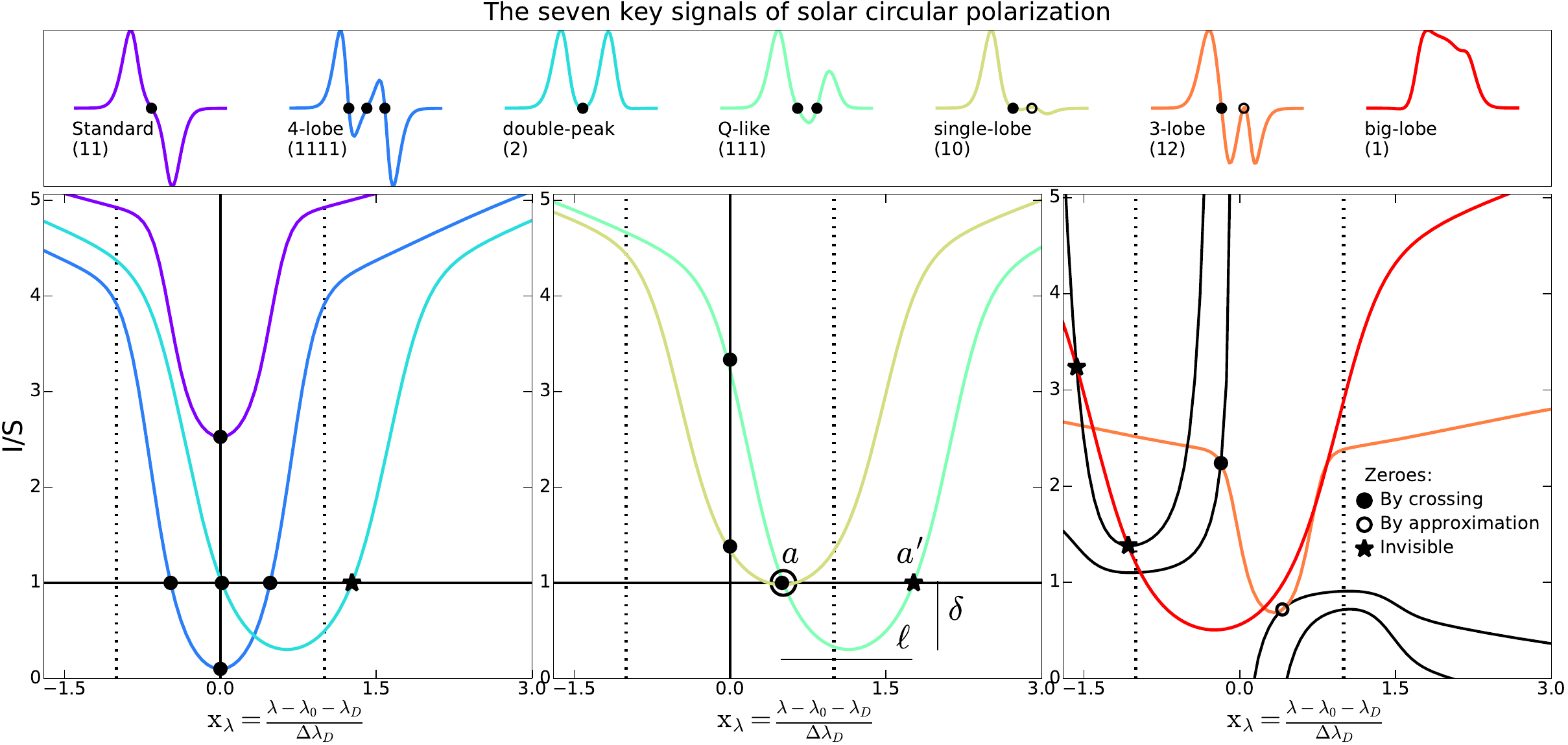}  
\caption{Representative solar Stokes V
  signals (upper panel) explained by associating
  their zeros (see legend) with intersections between background
  (colors) and critical (black)
  intensities. Vertical dotted lines limit the relevant
  absorption region ($|x_{\lambda}|<1$). The medium is weakly
  magnetized ($\beta = 0.05$),
  and optically thick ($T_{\nu_0}=3$) for all cases, but
  $T_{\nu_0}=1.5$ for the three-lobe profile. The values of
  continuum-to-line opacity ($r_c=0.01$) and atomic orientation
($w^1_0< 0$ in the right panel or $w^1_0=0$ in the other ones) were chosen to ease the explanations.}
\label{fig:fig14}
\end{figure*}

\section{Zeroes-based morphology of Stokes V in a single magnetic layer}\label{sec:zero-based}
In this section we use our model to develop a way of
understanding the polarization by means of its spectral zeroes. 
A meaningful approach must start
by the case of maximum resolution (in space, time, and in LOS depth),
which implies considering that the whole formation region can be
represented by a depth-resolved (i.e., single) magnetic layer ($V_0=0$). Future
developments could then model the spatio-temporal structure of
the scattering object (a solar structure, a solar region, or a whole
star) to obtain its polarization fingerprint as a combination of the basic pieces explained here. We focus on the most common case of an
absorption spectral line in which the effective
Zeeman splitting is smaller than the width of the optical
coefficients, such that the Zeeman components are not fully 
resolved. From here, the cases of emission line and very strong field
could easily follow.

Unless the contrary is specified, the calculations in this section
 correspond to solutions of the SEE in the reference atomic
system of the Na {\sc i} D lines. In general, every 
atomic system has a different efficiency converting radiation
field orientation into atomic orientation, and that is why the SEE has
to be solved for every line of interest. 
In this sense, our approach in this section is equivalent to a 
 sophisticated parametrization of the atomic
orientation through RF orientation, by means of a reference atomic
system that transform one into the other.
This approach still allows to extract general conclusions
about the polarization
signals of any spectral line because its morphology is
 ultimately due to the radiative transfer part of the
problem. Thus, one can first understand the connection
between the Stokes
signals and the radiative transfer quantities (generalized
variables), and then investigate separately the atomic particularities able to
produce (or not) those quantities.

\subsection{The critical intensity spectrum and the seven representatives of
solar circular polarization}
The left panel of
Fig. \ref{fig:fig12} shows that atomic orientation ($w_1$)
produces a spectrum of critical intensity with two roughly antisymmetric spectral
branches: a left and right one. If orientation increases, the critical spectrum bends
itself around attractor points at
$x_{\lambda}=(\lambda-\lambda_0-\lambda_D)/\Delta \lambda_D \approx
\pm \, 2$. On the contrary, if $w_1\rightarrow 0$
the two branches join at right angles where $I=S$ and $x_{\lambda}=0$. The
vertical line at $x_{\lambda}=0$ represents a jump (a discontinuity) between $1$ and
$\infty$ at the blue spectral side, and between $-\infty$ and $1$ at the
red side, and must be considered for counting the crossings of
Stokes V through zero. Some deviations from this general pattern can
be further studied by varying the opacities.

 The spectra of critical intensities allow a better understanding of the
morphology of polarization signals because the number of times that the
illuminating intensity $I_0$ crosses a given critical intensity curve determines the
number of zeroes and peaks that
Stokes V has, hence its shape. There being $M$ total wavelengths
where $I \approx I^{\rm critic}_0$, there are $N$ crosses in the
wavelength span of the absorption coefficient (``inner'' zeroes) 
corresponding with $N$ visible zeros and $N+1$
peaks in Stokes V, with the other $M-N$ zeros now being invisible (``outer''
zeros). Sometimes the intensity at a given
 wavelength does not cross the critical value, but approaches it enough
 to cancel the V signal significantly. Thus, we shall then say that,
 among the N inner zeros, some
 of them can be considered as \textit{zeros by proximity}, not by crossing. 
  This simple analysis
 allows one to quickly classify and understand any polarization
 signal. Figure \ref{fig:fig14} and Table \ref{tab:morph}
 show this for the seven circular polarization signals
that the author considers the most representative, based on the solar
observations cited in the second paragraph of the introduction
and on their theoretical interest for solar diagnosis. 

\begin{table}[h!]
\caption{Morphological classification. Names in first column distinguish
    alternate-sign signals from single-sign ones: e.g., a
    three-lobe signal differs from a triple-lobe (not
    shown) in that the latter has all peaks of the same
    sign. Names in second column identify signals by the number of
    consecutive peaks with the same sign: e.g., $12$ (one, two)
  means one peak of one sign and two peaks of the opposite sign. $N_{\rm in}$ is number of
    zeroes in absorption band and $N_{\rm ap}$ how many of them
are ``by approximation''.}
\centering %
\begin{tabular}{|c|c|c|c|c|c|}
\hline
\begin{tabular}{@{}c@{}} Name \end{tabular}  & 
\begin{tabular}{@{}c@{}}Precise \\ name \end{tabular}  &
\begin{tabular}{@{}c@{}@{}}$N_{\rm in}$/ \\$N_{\rm ap}$\end{tabular} &
$\delta$ &
\begin{tabular}{@{}@{}c@{}@{}}Not \\possible \\ if \\ $\rho^1_0=0$\end{tabular} & 
\begin{tabular}{@{}c@{}}Emit \\ $J^1_0$\,$\uparrow$ \end{tabular}\\
\hline
Standard & $11$ & $1/0$ & $+$ & - & -\\
Four-lobe & $1111$ & $3/0$& $-$ & - & -\\
Double-peak & $2$ & $1/0$ & $-$ & - & \checkmark\\
Q-like & $111$ & $2/0$  & $-$& -  &  \checkmark \\
Single-lobe & $10$, $01$ & $2/1$  & $0$& - &  \checkmark \\
Three-lobe & $12$, $21$ & $2/1$ & $-$ & - & $\approx$\\
Big-lobe & $1$ & $0/0$  & $-$&  \checkmark   &  \checkmark \\
\hline
\end{tabular}
\label{tab:morph}
\end{table}
The obtention of  
Fig. \ref{fig:fig14} is direct once it is realized that every kind of
Stokes V signal forms in specific bins of Doppler
velocity (gradient) and $I_0$. The bins in velocity are
centered on particular values of $\xi$ that shifts the characteristic
points of the absorption profile\footnote{The relevant points are: the line
center, where $I_0$ is minimum; the crosspoints between $I_0$ and
$I^{\rm critic}_0$, where the medium is neutral; 
and the crosses at the half-width of the profile, where the spectral gradient of
$I_0$ is maximum.} of $I_0$ to the
characteristic points of $\eta_V$ ($x_{\lambda}=0$ and
$x_{\lambda}\approx 1/2$),
thus giving a total of $20$ values of $\xi$. Studying all of them we could
explore the relative variations of amplitude between the peaks of the
signal, but the morphological essence (number of peaks and zeroes) is
 only determined by the points $a$ and $a^{\prime}$ marking the wavelengths of neutral medium. 
This realization leads us to consider only $\xi_0=0$ and three characteristic Doppler shifts (plus their
symmetric ones) linking velocity and width: 
 $\xi_1=1-\ell/2$, $\xi_2=\ell/2$, and
$\xi_3=(1+\ell)/2$. The distance $\ell=|x_a-x_{a^{\prime}}|=p\cdot w$ (with $p$ a given fraction of the $I_0$
width $w$) is generically
plotted in the middle panel of Fig. \ref{fig:fig14}. Below, we explain
the profiles in the figure:
\begin{itemize}

\item The \textbf{standard} antisymmetric 
Stokes V signals (and their near-antisymmetric relatives when
atomic orientation and/or an imbalance of illumination exists) result
from a single cross in the spectral width of the absorption
profile that happens 
when\footnote{Note that regardless of
  the amount of orientation, the
  intensities that are above the line source function can cross the critical
  intensity once.} $I_0>S$ ($\delta>0$). Any other signal is considered
nonstandard and occurs for $\delta\leq 0$.
\item \textbf{Four-lobe Stokes V profiles} (i.e., three inner zeroes and four
peaks) are formed when the $I_0$ core drops below
the source function value of the scattering layer with $\ell\lesssim 3/2$ and 
Doppler shift $\xi \lesssim \xi_1$. The inner peaks are enhanced with a weaker $I_0$. The inner peaks of 
similar profiles have been
associated with magneto-optical effects in \cite{LL04} and also
with the relative spectral location and strength of the Zeeman sigma
components. But we 
see here that they are easily produced by a mere background
intensity with $\delta<0$. Hence, we think those previous explanations are
incomplete, if not incorrect. 

\item A \textbf{double-peak profile} (i.e., only two
  peaks of whichever relative amplitude and width but with
  the same sign) is exactly formed with a single inner zero that is furthermore
  at the center of the
absorption profile. As deduced from the lower-left panel of Fig. \ref{fig:fig14}, this
 requires: 1) a velocity gradient shifting the inner zero
 \textit{a} to $ x_a=0$ (i.e., $\xi =\xi_2=\ell/2$); and
2) an $I_0$ broadening large enough to
shift the zero $ a^{\prime}$ out of the absorption band ($x_{a^{\prime}}>1$). 
Substituting $\ell$, these conditions imply 
 $\xi/p= 0.5$ and a minimum $I_0$ width such that $p\cdot\alpha\geq
 1$. These conditions can be written in terms of $\delta$ if a ligature
 $p=f(\delta,\alpha)$ is set by specifying an $I_0$ profile. 

It seems incompatible to have atomic orientation and a
{perfectly symmetric} double-peak profile at the same solar location
because orientation moves the inner zero away from $x_{\lambda}=0$
regardless of the shape of $I_0$. However, irradiating double-peak Stokes V
signals from a solar location introduces orientation in the atoms
of the surroundings. Hence, anomalous signals (especially double peaks)
 may reveal the presence of atomic orientation around their locations.

\item A \textbf{Q-like Stokes V profile} (three peaks of
alternating signs) is created when there is
  a single inner zero per branch. 
For this, $I_0$ must be deep enough to intersect the two branches of
$I^{\rm critic}_0$,
 but also broad and shifted enough to intersect each branch only once in $x_{\lambda}\in
[-1,1]$. This implies $\xi_2<\xi<\xi_1$ with $\ell<0.5$. The
Q-like profile in this figure has a central peak that cannot be as
large as the others: its amplitude is
restricted because without orientation, intensity cannot develop
 significant excursions between the two zeroes (as it did in the rightmost column
 of Fig. 12). The case without orientation has been chosen to simplify
 the Fig. \ref{fig:fig14}.

\item A \textbf{single-lobe V profile} is a particular case in which a zero by
  approximation takes place near $x_a=x_{a^{\prime}} \approx 1/2$
(i.e., $\delta\approx 0$ in case of $w_1=0$) with $\alpha \geq 1$,
which suppresses one lobe. Atomic
orientation alters $\delta$, making suppression difficult. As happened with double-peak profiles, this inverse relation with orientation is
counter-intuitive because single-lobe profiles still imply a significant
increment of the irradiation of NCP to the surroundings, where signals affected by atomic
orientation should then be expected.

\item Figure\ref{fig:fig14} indicates that \textbf{three-lobe Stokes V} signals
produced by a single magnetic layer 
require one zero by crossing, one zero by approximation, and $\xi
\approx 1/2$, but now the
FWHM of I$_0$ must be narrower than half of the
absorption band ($\alpha<1$). Despite the fact that the calculations indicate that three-peak
profiles can be produced without orientation, the relative amplitudes
between the peaks (Fig.~\ref{fig:fig16}) seem to require orientation
(pending confirmation).

\end{itemize}

\subsection{Stokes V zeroes across the space of parameters: atomic
  polarity and the laws of proportionality.}
By exploring the variations of circular
polarization with the parameters of our model,
we can point out some symmetry properties and reconsider the role of
atomic orientation in weak and strong fields. 

\begin{figure*}[t!]
\centering$
\begin{array}{cc}
\includegraphics[scale=0.49]{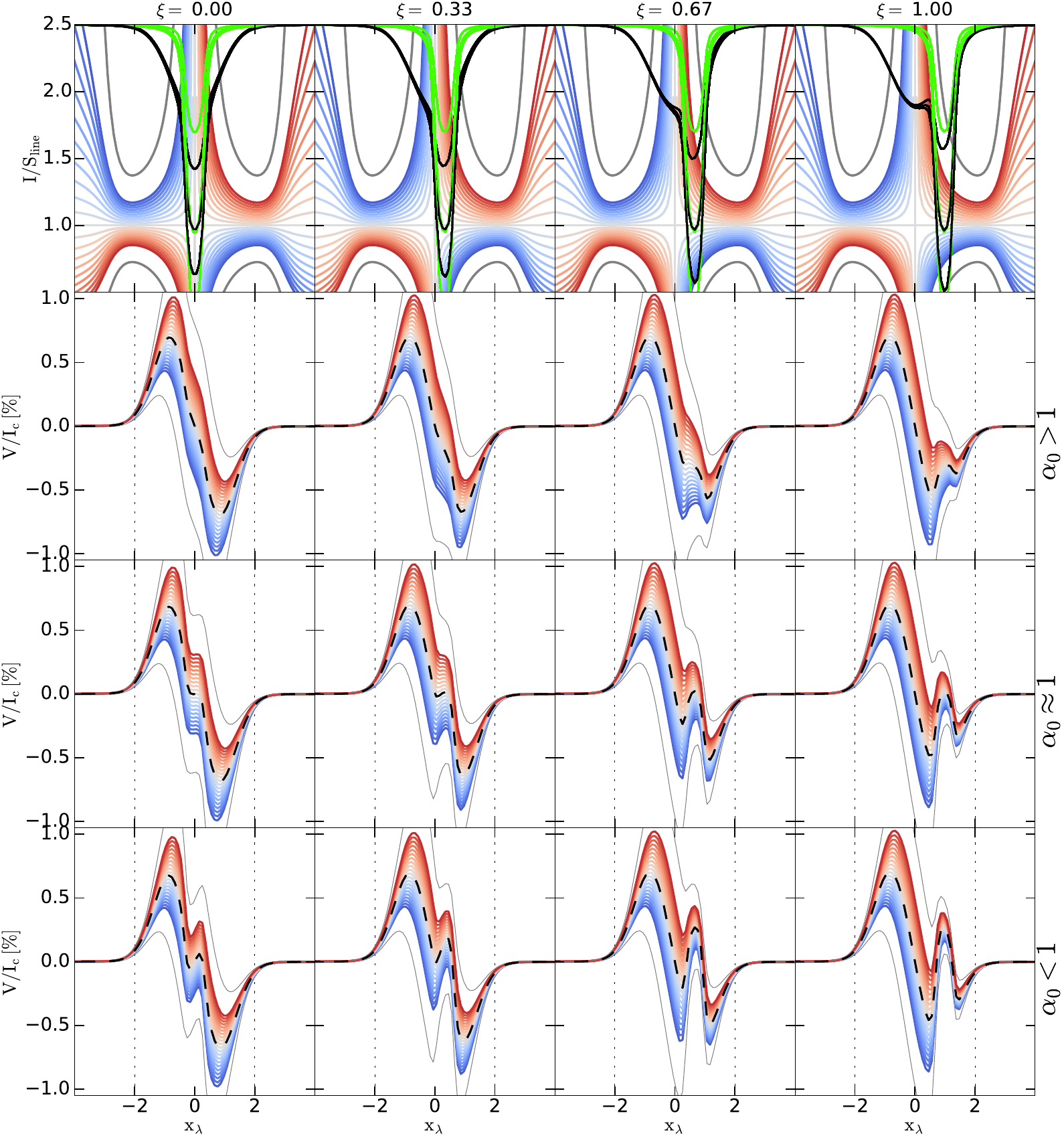} &    
\includegraphics[scale=0.49]{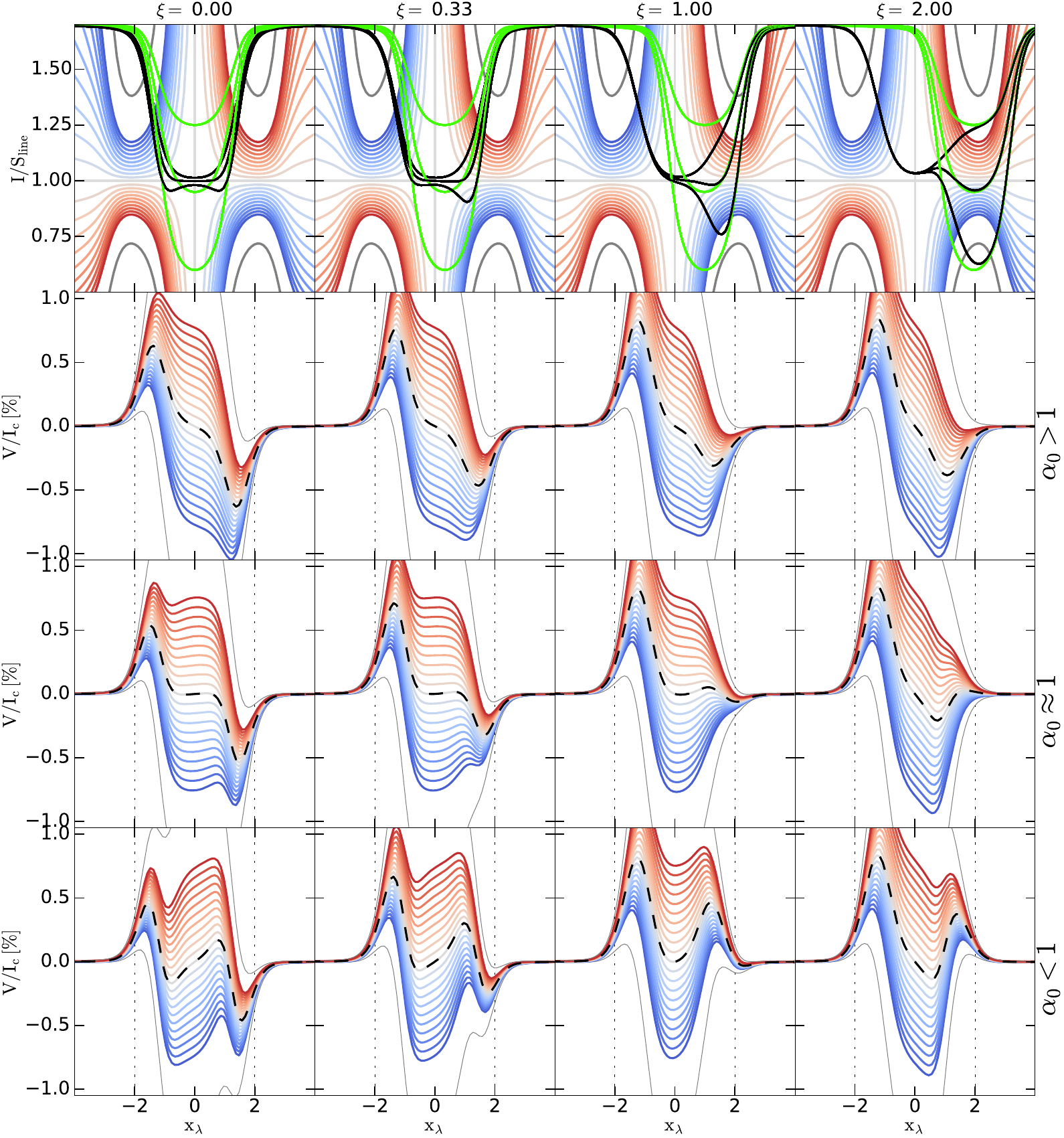}\\    
\includegraphics[scale=0.49]{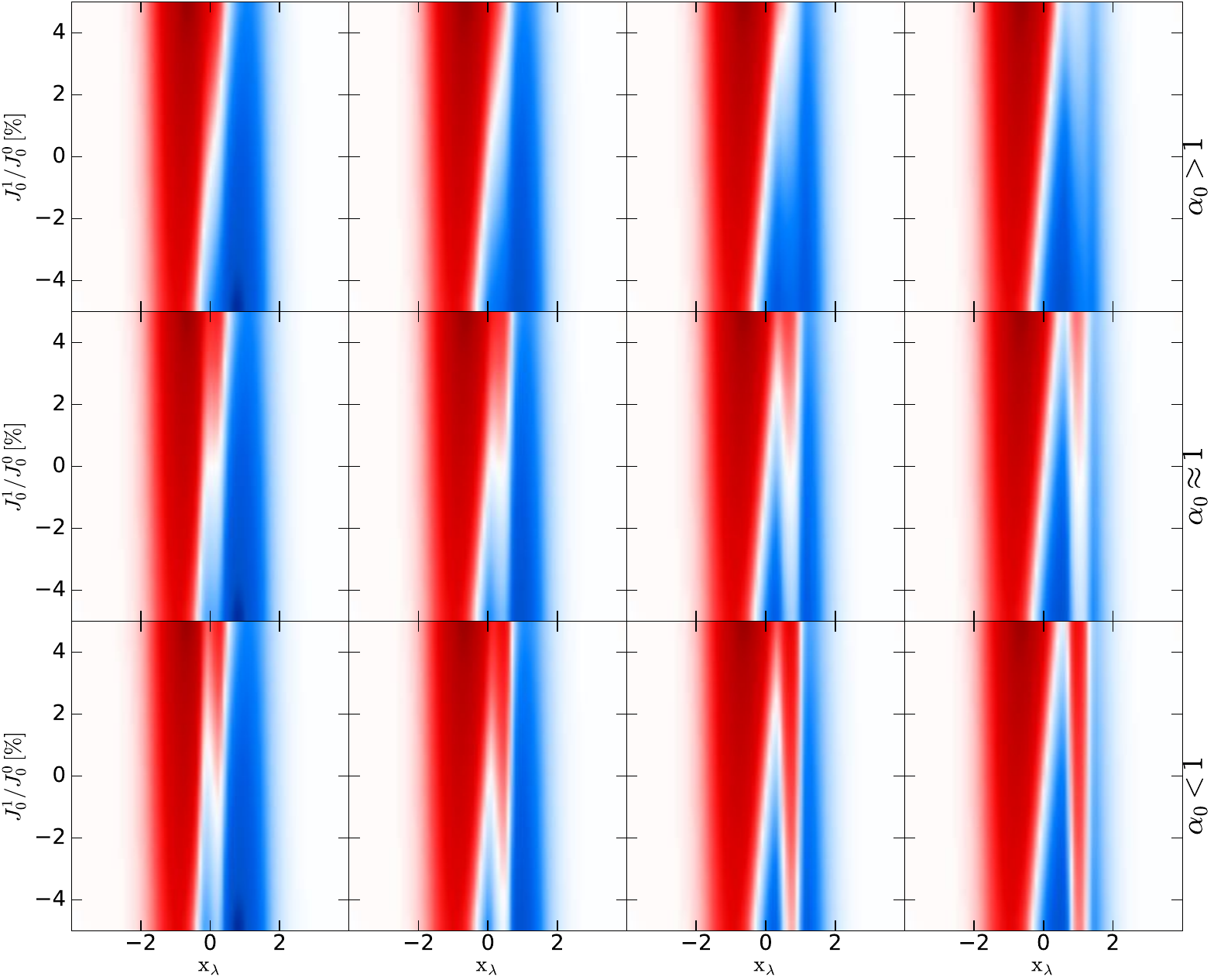}&
\includegraphics[scale=0.49]{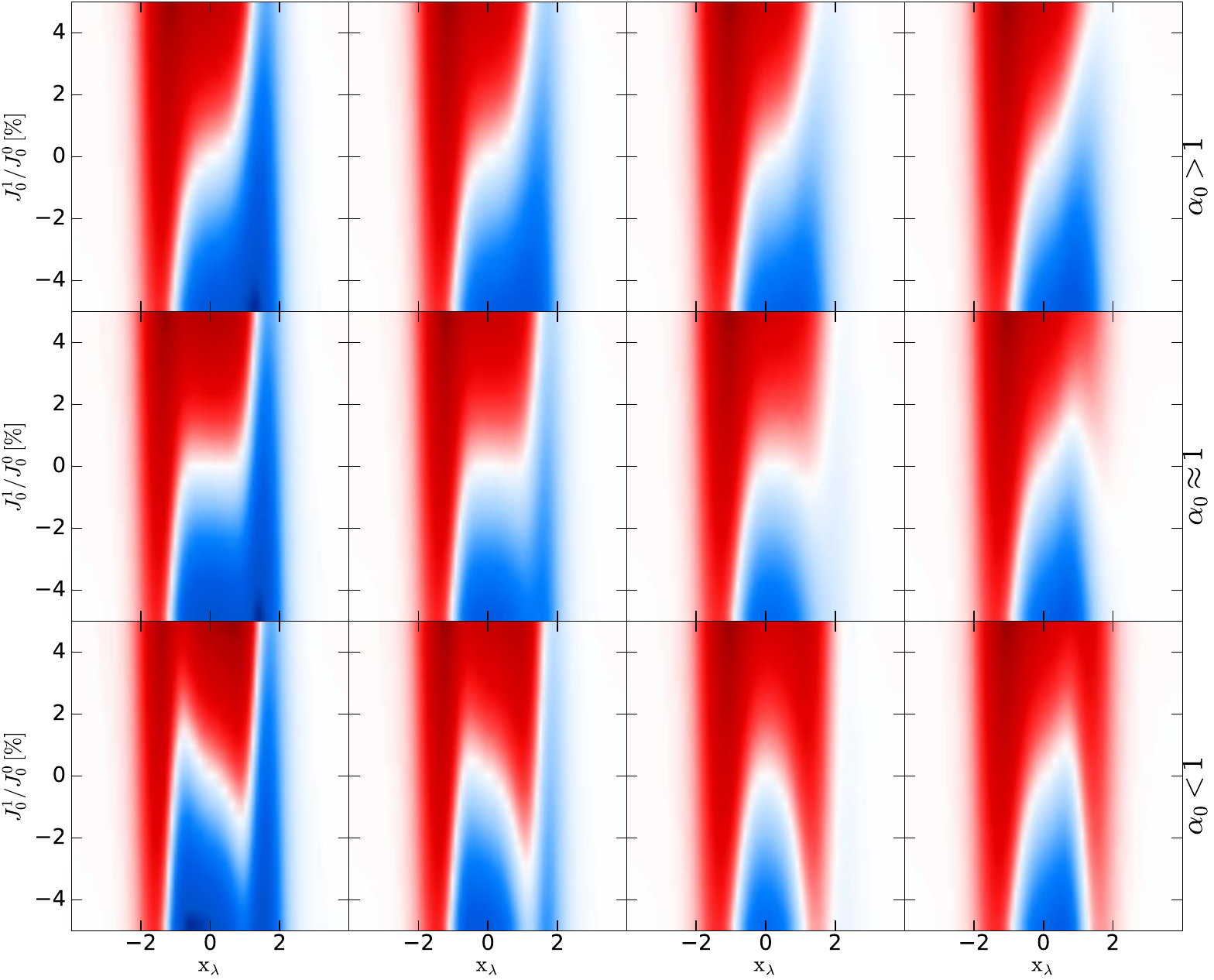}  
 \end{array}$
\caption{ Similar to Fig. \ref{fig:fig12} but varying $\alpha$,
  $T_{\nu_0}$, $\xi$ (by columns), and $I_0$ (with
  $\alpha_0=I_0/S_{\rm line}$ decreasing by rows). Left upper panel:
  $V/I_c$ for $T_{\nu_0}=0.5$ and $\alpha=1/3$. Right upper panel: As in the left panel but for $T_{\nu_0}=3$ and
  $\alpha=1.4$. Lower panels: As in upper panels but in two dimensions. }
\label{fig:fig13}
\end{figure*}

In the upper panels of Fig. \ref{fig:fig13}, the
spectral variations of Stokes V are plotted for strategically sampled 
values of $\alpha$, $\xi$, $w_1$, and
$T_{\nu_0}$. When considering the continuous changes necessary to
connect the different plots, 
these figures account for many of the 
circular polarization signals that an
absorption spectral line can produce under solar
conditions with $\beta<0.5$. 

The lower panels of Fig. \ref{fig:fig13} highlight in white color the evolution of
the zeroes of the circular polarization: anomalous signals appear when
there are two or more zeroes in the absorption band.  
As a side note, we have noted that Stokes V seems to follow a sort of
self-similar relationship such that its morphology is preserved when
opacity, $\alpha$, and $I_0$ change together with a nonidentified proportion among
them. 
\begin{figure*}[h!]
\centering
\includegraphics[width=0.79\linewidth]{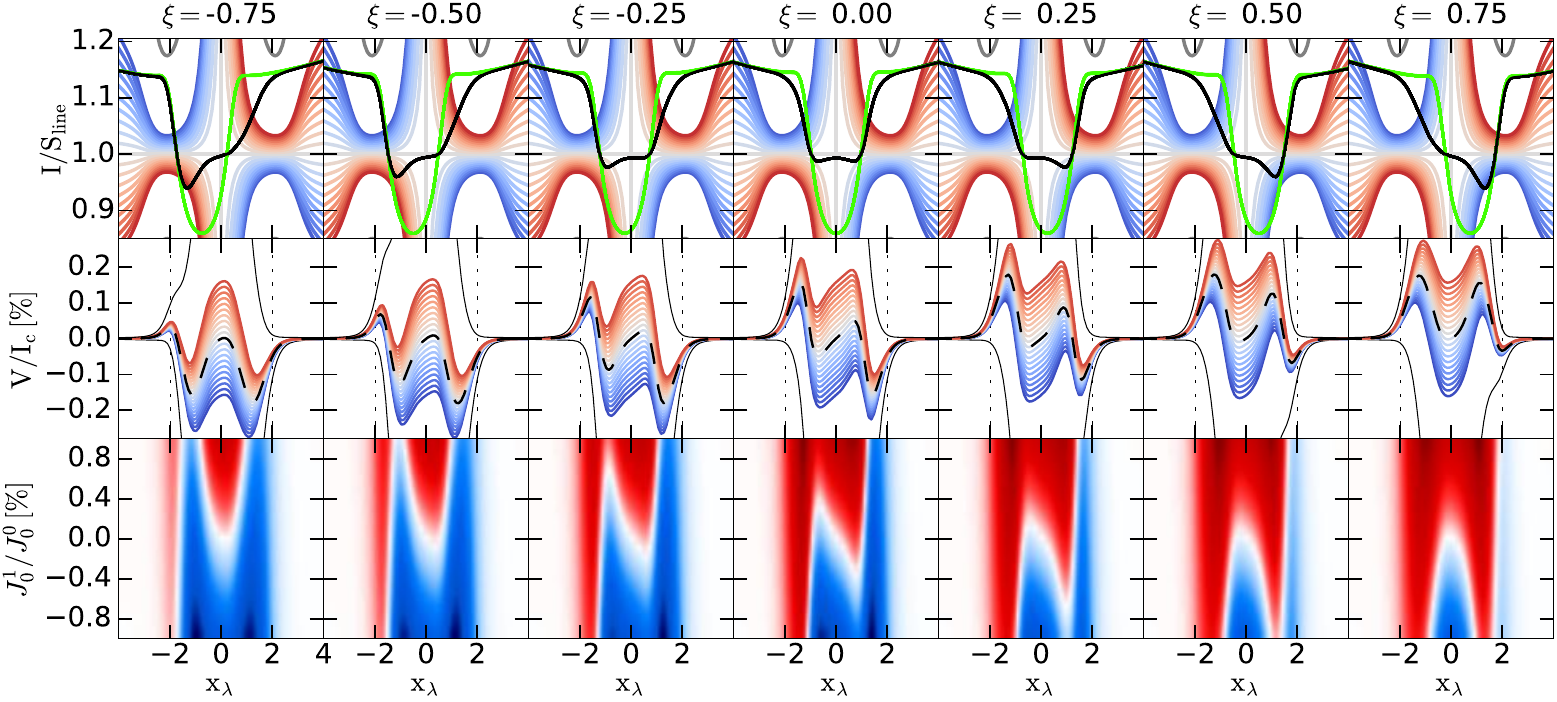}
\caption{As in Fig. \ref{fig:fig13} but with $w^1_0=1\,\%$,
  $\alpha=1.2$, reduced $I_0$ (see axes scale) and including cases with $\xi<0$.}
\label{fig:fig15}
\end{figure*}
Let us now focus for a moment on the particular case of a perfectly symmetric double-peak
profile, which must have a
single inner zero located at $x_{\lambda}=0$ when $w_1=0$.
If atomic orientation is added, the two branches of critical
intensity detach away from their junction at $x_{\lambda}=0$ and
the direction of separation depends on the sign of the atomic
orientation. Considering now a modification of velocity gradient
 $\xi$ from the value that produces the double-peak profile, it can
  be seen (with help of Figs.~\ref{fig:fig12} and \ref{fig:fig14})
  that for certain values, the detachment due to orientation leads to
  two inner zeroes instead of one when $w^1_0 \gtrless 0$ and
$\xi \lessgtr 0$ (i.e., when velocity gradient and orientation have opposite signs), or no
zeroes at all when $w^1_0 \gtrless 0$ and $\xi \gtrless 0$ (i.e., when
they have same sign). This reveals two interesting facts. One is that
particular combinations of signs of orientation and velocity
gradient are linked to different {morphologies} in Stokes V. The other is that the ways to
explain a given transformation towards a double-peak profile are
different depending on the signs of atomic orientation and velocity. In other words,
one can always approach a given Stokes V profile from completely different
scenarios in the space of parameters. As shown in Fig. \ref{fig:fig13}, this actually holds for any
other kind of Stokes V signal, not only for double-peak profiles.

All this is better seen with Fig. \ref{fig:fig15}, when the
(redundant) profiles for $\xi<0$ are included. Then, the following symmetry
property for velocity gradient
and radiation field orientation is identified:
\begin{equation}
V(\xi,w_1,x_{\lambda})=-V(-\xi,-w_1,-x_{\lambda}),
\label{eq:symmetry}
\end{equation}
We note that the effect of velocity gradients of opposite signs is to reverse the red and
blue part of the profiles ($x_{\lambda} \leftrightarrow
-x_{\lambda}$) at the same
time as inverting the sign of the signals. Hence, as shown by Fig. \ref{fig:fig15}, all signals can
be unequivocally associated to a unique configuration of orientation and velocity
gradient (no ambiguity). However, when considering the direction
 $\vec{\Omega}_B$ of the magnetic field:
\begin{equation}
\centering
V(\vec{\Omega}_B,\xi,w_1,x_{\lambda})= V(-\vec{\Omega}_B,-\xi,-w_1,-x_{\lambda}).
\label{eq:symmetryB}
\end{equation}
Now, there is ambiguity (in particular, in the polarity of the
magnetic field) 
when the signals are symmetric around line
center. Asymmetric signals can avoid the ambiguity because nothing
reverses the profile spectrally without inverting its signs, and because the
observer can always distinguish the red and blue halves of a profile. Interestingly, atomic
orientation is a source of asymmetry, hence it can contribute to avoiding ambiguity.

\subsection{Atomic polarity}
Figures \ref{fig:fig13} and \ref{fig:fig15} show that there is a third
kind of polarity, that introduced by atomic orientation. When the
effect of the orientation overcomes that of the magnetic field, all
profiles adopt a definite sign corresponding to the sign of
orientation. Therefore, at any
given wavelength the balance between the three kinds of polarity (magnetic, radiative, and
atomic) determines the real polarity of Stokes V. Hence, we realize
that in general the 
combination of these factors causes the concept of polarity for 
Stokes V
 to lose its meaning, because the three kinds of sign drivers
 are in principle uncorrelated and can change along the
LOS, producing an
arbitrary combination of signs in a given profile. Indeed, that could be the
very definition of an anomalous Stokes V signal: one reflecting a
combination of polarities (of whichever kind).

To see how difficult  the association becomes with a
  magnetic polarity in a general situation, we can try 
to define a nonambiguous reference polarity:

\textbf{\textit{Reference polarity} ($P_r$)}: assuming that a signal
is fully resolved in space and time and that the longitudinal magnetic
field does not change sign along the LOS, the reference
polarity of an antisymmetric Stokes V profile corresponding to
a spectral line with $g_{\rm eff}>0$ and without atomic polarization 
is defined as positive when the
longitudinal magnetic field at the scattering layer 
points towards the
observer (positive magnetic polarity) and
$I_0/S>1$ (positive radiative polarity).\\

\subsection{Laws of proportionality: is atomic orientation
  relevant when the magnetic field is not weak?}
The expected physical situation 
 is that atomic orientation is {only} important in media with weak
  magnetic fields, but it is worthwhile investigating some arguments that
  could challenge this viewpoint. In this paper we start to develop this idea by introducing
 the following laws of proportionality:
\begin{itemize}
\item Law 1: for producing the
same relative morphological change in Stokes V, the more atomic
orientation is needed the larger the magnetic
field. This was shown in Fig. \ref{fig:fig0}.
\item Law 2: the larger the NCP of a pumping Stokes V radiation
    field, the more atomic orientation
    is induced in the pumped scatterers. This is a
natural consequence of Eqs. (\ref{eq:j10j00}) and
(\ref{eq:ncp_ed}).  
\item Law 3: the larger the magnetic fields are in the
    surroundings of the scatterer, the larger the Stokes V signals are that
    pump it, simply because the Zeeman components cancel each
 other less efficiently (they are more separated). As a rule of
    thumb, Stokes V in hecto-Gauss fields is often an order of magnitude
larger than in weak fields.
\item Law 4: if the Stokes V signals
    pumping the scatterer 
    are anomalous: the larger their
    amplitudes, the larger the NCP (hence the orientation
    created). This is especially relevant in double-peak profiles
    because they only have one sign.
\end{itemize}
The combination of Laws 3 and 4 with the fact that
NCP can be an order of magnitude larger in anomalous signals than in
standard signals (see
Sect. \ref{sec:ori2peaks}), 
implies that larger magnetic fields
should induce proportionally larger atomic orientation. If the
proportionality factor is large
 enough (and this is the key), 
Law 1 says that similar effects as in weak field should be observed. 
This argument is reinforced by the results 
  presented in previous sections, which show that
 the creation of anomalous signals with enhanced NCP does not 
critically depend on the magnetic field strength, but on the balance
between emission and dichroism. Indeed, the possibility of partially changing
the Stokes V polarity radiatively implies that, independently
of the field strength, similar dichroic sign inversions can also be induced along the
other rays of light pumping a scatterer from all directions inside
the atmosphere. This seems to be supported by the fact 
that anomalous CP is observed all over the disk (see observational
bibliography in introduction), but also at the limb, as indicated
by our measurements with ZIMPOL.

Hence, although caution is advised, 
we should investigate whether it is possible to
find situations where both atomic orientation and {}hecto-Gauss fields shape
the polarization. Two questions arise: Is a three-dimensional
optical pumping able to produce visible orientation effects in hecto-Gauss
fields? What is the maximum magnetic field 
strength that allows observation of the signature of orientation? This
could be investigated if we confirm our finding about the existence
 of a stable point where the Stokes V amplitude depends
 robustly on atomic orientation for any background illumination (Sect. \ref{detail}).

\section{Conclusions}
By developing a comprehensive two-layer radiative-transfer model, 
we studied the NLTE generation of anomalous Stokes V
signals with magnetic and nonmagnetic
 dichroism. We
considered a 
first physical scenario able to produce 
double-peak and Q-like Stokes V signals without longitudinal magnetic field (no Zeeman splitting). This effect
consists in the self-absorption of the central part
of a polarization profile with only one lobe due to the action of
 (orientation-dominated) dichroism. Its explanation served to characterize
and understand the formation of dichroic polarization from an academic
perspective and to introduce
useful concepts such as neutral medium, reinforcing
medium, critical intensity spectra, and critical source function. Thus, this first
version of the model reveals the importance of 
the balance between polarized emission and dichroism. 
A remarkable finding was that the signs of the ratio
$\epsilon_V/\eta_V$ and the level of intensity cause
dichroism to act on polarization as an amplifying or nullifying mechanism. 

Using these concepts we also studied a second scenario based on the
uneven pumping of Zeeman components when both magnetic and
nonmagnetic dichroism act together. 
Thus, we identified the basic conditions explaining anomalous
Stokes V signals for a generic
solar spectral line.
The first key is the combination of: 1) dichroism,
which allows intensity to
modify the polarization; 2) atomic
polarization, a form of dichroism that modifies the shape of the optical coefficients and
the sensitivity of the atomic system to intensity-driven sign
reversals; 3) velocity gradients, which restrict or enhance the action of
dichroism at particular wavelengths; and 4) a magnetic field, 
which naturally allows non-null optical coefficients in cases without atomic
polarization and allows V profiles with more than one
lobe to be explained. 

The second key to explaining polarization anomalies has to do with
the resolution of the physics along the LOS, namely: 1) the particular 
behavior of the source function for intensity, which can produce a 
radiative sign inversion of a whole polarization profile; and 2) the need for more
than one magnetic component along the LOS to explain all the
features of the observed
profiles  (shape, peak separation, and amplitude). To show this, we
used our model to
reproduce anomalous double-peak and Q-like profiles such as those observed
in Na {\sc i} D$_1$ and D$_2$, and we presented a preliminary 
modeling of the observed anomalous CP in the Fe {\sc i}
$1564.85$ nm line without the need to assume unresolved 
magnetic field components. 
We advance that our model allows to
reproduce the anomalous CP and intensity
measured in the solar context of umbral flashes
(\cite{Socas-Navarro:2000aa}), again without assuming lack of 
resolution (paper in prep.). 

In general, these results warn against the
common assumption of spatially unresolved magnetic
components which, justified or not, tends to ignore the origin of the
polarization by overlooking the
physics of polarized radiative transfer explained here.
This emphasizes the need to 
maximize the spatiotemporal resolution, and the need to 
verify when a structure is or is not resolved, in order to 
understand the morphology of polarization and the solar atmosphere. 


We also developed an insightful way to explain
 the polarization morphology, based on the
zeroes of the spectral profiles. The so-called ``inner zeroes''
are progressively shifted to the line wings and transformed into
invisible ``outer'' zeroes as an increasing velocity gradient
between background and scattering layers shifts the minimum
background intensity towards 
wavelengths with insignificant absorption. Thus, the
explanation and classification of the seven most representative Stokes V
solar profiles  followed
easily. They were classified as standard (or $11$),
single-lobe ($10$ or $01$),
big-lobe (or $1$), double-lobe (or $2$), Q-like (or $111$), three-lobe ($12$ or
$12$), or four-lobe (or $1111$). All
anomalous signals are produced in conditions of low relative
intensity. 
The morphology of anomalous polarization signals is very sensitive 
to the line-broadening mechanisms, to dynamics, and to the levels of
intensity inside the atmosphere,
and for this reason they seem very valuable for testing and improving
MHD models. Anomalous CP is easy to find in dynamic
simulations and in observations.

We have shown
that the polarity of a resolved Stokes V profile 
is a mixture of magnetic, radiative, and atomic polarities, the
anomalous signals occurring when the spectrum is not dominated by only
one of them. Regarding atomic orientation, we explained how it modifies
Stokes V, emphasizing a law of proportionality between
orientation and field strength that is reinforced in the presence of
anomalous CP, and that suggests that atomic orientation could play a role in nonweak
magnetic fields. Future investigations are needed to confirm whether
 solar atomic orientation can be effectively measured by
identifying in the profile of Stokes V a ``robust point'' that is
stable against changes in background illumination. 
Finally, our calculations led to the identification of the spectral symmetry
relationships relating velocity gradients, atomic orientation,
and magnetic field. Asymmetric signals influenced by atomic orientation
could be a good candidate for
avoiding magnetic field ambiguities. More investigation is
necessary in this direction.  



Several studies complement this paper. Of particular
interest are new observations of anomalous solar circular
polarization, the application of our model to LP, and a more detailed
study of the optical coefficients modifying the dichroic response of the scatterers.

\begin{acknowledgements}
      The author thanks L. Belluzzi and M. Bianda for their kind
      supervision and for useful scientific conversations during the initial stages of this investigation. He
is also grateful to J.M. Borrero, for sharing his observations
and the details of his investigation, and to N. Bello-González for
indirectly helping to orient the results of this paper. 
This investigation has been partially financed by the Swiss National Science
Foundation project $200021$\texttt{\_}$163405$ and the Spanish Ministry of Economy and Competitiveness (MINECO) under the Severo Ochoa Program MINECO SEV-2015-0548.
\end{acknowledgements}

%
%


\end{document}